\documentclass[pdflatex,sn-mathphys]{sn-jnl0}
\usepackage{graphicx}

\usepackage{amssymb,amsthm,amsmath}
\usepackage{xcolor,paralist,hyperref,titlesec,fancyhdr,etoolbox}
\usepackage{float}
\usepackage{xcolor}
\usepackage{subfiles}
\usepackage{dcolumn}% Align table columns on decimal point
\usepackage{bm}% bold math
\usepackage{verbatim}
\usepackage{upgreek}
\usepackage{etoc}
\newcommand{\cutwoo}{$\mathrm{Cu_2O}$ }

\begin{document}

%TC:ignore

\title[]{Nonlinear Rydberg exciton-polaritons in \cutwoo microcavities}


\author[1]{\fnm{Maxim} \sur{Makhonin}}%\email{m.makhonin@sheffield.ac.uk}
\author[1]{\fnm{Anthonin} \sur{Delphan}}%\email{adelphan1@sheffield.ac.uk}
\author[2]{\fnm{Kok Wee} \sur{Song}}%\email{k.song2@exeter.ac.uk}
\author[1]{\fnm{Paul} \sur{Walker}}%\email{p.m.walker@sheffield.ac.uk}
\author[1]{\fnm{Tommi} \sur{Isoniemi}}%\email{}
\author[1]{\fnm{Peter} \sur{Claronino}}%\email{}
\author[3]{\fnm{Konstantinos} \sur{Orfanakis}}
\author[3]{\fnm{Sai Kiran } \sur{Rajendran}}
\author[3]{\fnm{Hamid} \sur{Ohadi}}%\email{ho35@st-andrews.ac.uk}
\author[4]{\fnm{Julian} \sur{Heckötter}}%\email{julian.heckoetter@tu-dortmund.de}
\author[4]{\fnm{Marc} \sur{Assmann}}
\author[4]{\fnm{Manfred} \sur{Bayer}}%\email{manfred.bayer@tu-dortmund.de}
\author[1]{\fnm{Alexander} \sur{Tartakovskii}}
\author[1]{\fnm{Maurice} \sur{Skolnick}}%\email{}
\author[2]{\fnm{Oleksandr} \sur{Kyriienko}}%\email{O.Kyriienko@exeter.ac.uk}
\author[1]{\fnm{Dmitry} \sur{Krizhanovskii}}%\email{d.krizhanovskii@sheffield.ac.uk}

\affil[1]{\orgdiv{Department of Physics and Astronomy}, \orgname{University of Sheffield}, \orgaddress{\city{Sheffield}, \postcode{S3 7RH}, \country{UK}}}

\affil[2]{\orgdiv{Department of Physics and Astronomy}, \orgname{University of Exeter}, \orgaddress{\street{Stocker Rd}, \city{Exeter}, \postcode{EX4 4PY}, \country{UK}}}

\affil[3]{\orgdiv{SUPA, School of Physics and Astronomy}, \orgname{University of St Andrews}, \orgaddress{\city{St Andrews}, \postcode{KY16 9SS}, \country{UK}}}

\affil[4]{\orgdiv{Fakultät Physik}, \orgname{TU Dortmund}, \orgaddress{\street{August-Schmidt-Straße 4}, \city{Dortmund}, \postcode{44227}, \country{Germany}}}

%TC:endignore
%TC:break abstract
%%\newpage
%%\newpage

\abstract{

Rydberg excitons (analogues of Rydberg atoms in condensed matter systems) are highly excited bound electron-hole states with large Bohr radii. The interaction between them as well as exciton coupling to light may lead to strong optical nonlinearity, with applications in sensing and quantum information processing. Here, we achieve strong effective photon-photon interactions (Kerr-like optical nonlinearity) via the Rydberg blockade phenomenon and the hybridisation of excitons and photons forming polaritons in a Cu$_2$O-filled microresonators. Under pulsed resonant excitation polariton resonance frequencies are renormalised due to the reduction of the photon-exciton coupling with increasing exciton density. Theoretical analysis shows that the Rydberg blockade plays a major role in the experimentally observed scaling of the polariton nonlinearity coefficient as $\propto n^{4.4 \pm 1.8}$ for principal quantum numbers up to $n = 7$. Such high principal quantum numbers studied in a polariton system for the first time are essential for realisation of  high Rydberg optical nonlinearities, which  paves the way towards quantum optical applications and fundamental studies of strongly-correlated photonic (polaritonic) states  in a solid state system.

%Rydberg excitons (analogues of Rydberg atoms in semiconductors) are highly excited bound electron-hole states with large Bohr radii. 
%The interaction between them as well as exciton coupling to light may lead to strong optical nonlinearity, with applications in sensing and quantum information processing. 
%Here, we achieve strong effective photon-photon interactions (Kerr-like optical nonlinearity) via the Rydberg blockade phenomenon and the hybridisation of excitons and photons forming polaritons in a Cu$_2$O-filled microresonators.
%Here, we investigate Kerr-like nonlinearity for Rydberg exciton-polaritons (polaritons) in a Cu$_2$O-filled microcavity. 
%Under pulsed resonant excitation polariton resonance frequencies are renormalised due to the reduction of the photon-exciton coupling with increasing exciton density. Theoretical analysis shows that the Rydberg blockade plays a major role in the experimentally observed scaling of the polariton nonlinearity coefficient as $\propto n^{4.4 \pm 1.8}$ for principal quantum numbers up to $n=7$.  Such high exciton-polariton quantum numbers are not accessible in other polariton platforms. The strength of exciton-polariton nonlinearity is found to be comparable or even exceeding those in other highly nonlinear polariton platforms, such as GaAs or hybrid perovskites. Our work makes a fundamental step towards highly nonlinear Rydberg polaritonics paving the way to quantum optical applications and fundamental studies of strongly-correlated photonic (polaritonic) states in solids.
}


\maketitle
%%TC:break maintext
%%%
%%\begin{refsection}

%%%
\section*{Introduction}
Prior to the study of Rydberg excitons in solids,  significant efforts have been devoted to the research of Rydberg atoms --- giant atomic states with valence electrons occupying orbits of high energy excited states (with sizes up to tens of micrometers). Rydberg states have been at the focus of fundamental and applied science in areas of metrology~\cite{DingNatPhys2022}, sensing~\cite{Fancher2021,Fancher2022}, quantum information and simulation \cite{GallgherBook2005,GrahamPRL2019,Tiarks2019}. Their strong long-range dipole-dipole interactions lead to the Rydberg blockade phenomenon \cite{LukinPRL2001, Urban2009, Gaetan2009}, where the presence of one excited atom prevents the excitation of another in its vicinity, at the same frequency. This effect forms the basis for quantum information processing (QIP) with Rydberg atoms~\cite{Saffman_RMP2010,Wu_2021,Morgado2021, Shi2022}. Furthermore, coupling light to Rydberg atoms \cite{MurrayAdv2016,Fleischhauer2005} enables strong effective photon-photon interactions down to the single particle level paving the way towards the development of various quantum optical devices (single photon switches, phase shifters, transistors etc \cite{Firstenberg_2016}).

Recently, Rydberg excitons (the analogue of Rydberg atoms in condensed matter systems) were observed in a number of materials including transition metal dichalcogenides (TMDCs) \cite{Gu2021}, perovskites~\cite{Bao2019}, and \cutwoo{} \cite{Kazimierczuk2014,Amann2020,Morin2022,Steinhauer2020}, where the %Bohr
\textcolor{black}{Rydberg exciton} radius reaches a few microns in states with principal quantum numbers as high as $n = 30$~\cite{Versteegh2021} and \textcolor{black}{ultra-high nonlinearities have been reported in CW regime \cite{Morin2022,WaltherPRL2020}.} Rydberg exciton blockade was demonstrated experimentally ~\cite{Hecktter2021}. On the other hand, there is a strong interest in the study of hybridised excitons and photons in microcavities and waveguides, which lead to the realisation of giant Kerr-like optical (polaritonic) nonlinearities~\cite{CarusottoCiuti2013,Byrnes2014}. These nonlinearities can be exploited for development of highly nonlinear and scalable optical devices on a chip with possible applications in QIP~\cite{Kuriakose2022}. Polariton nonlinear phenomena, such as superfluidity, solitons, photon blockade~\cite{Delteil2019,MuozMatutano2019} and single-photon phase shifts~\cite{Kuriakose2022}, to name a few, were investigated mostly for 1s Wannier-Mott excitons. Nonlinear energy shifts have been reported for Rydberg exciton-polaritons in TMDCs~\cite{Gu2021,Chernikov2014, Coriolano2022} and perovskites~\cite{Bao2019}, but these studies are limited to the first two excited exciton states. By contrast, large principal numbers $n$ are available in \cutwoo{} \textcolor{black}{and very recently strong exciton-photon coupling was observed in a microcavity with embedded \cutwoo \cite{Orfanakis2022}, where only linear optical response of Rydberg exciton-polaritons was addressed.} %have potential to open a route to ultra-high polaritonic nonlinearity.

%%%
Here, we study yellow series Rydberg exciton-polaritons in a planar Fabry-Perot microcavity with embedded \cutwoo{} thin crystal. We report a strong and ultra-fast nonlinear optical response for these states and demonstrate the highly superlinear scaling of the nonlinearity with the principal quantum number up to $n=7$. This scaling opens up potential for ultra-high nonlinearity at the highest numbers ($n=30$) observed so far in bare \cutwoo{} crystals. 
The nonlinearities are found to be comparable or even exceeding (for $n\geq5$) the giant optical Kerr-like nonlinearities observed in other polariton platforms, such as GaAs or hybrid perovskites~\cite{Fieramosca_2019}. \textcolor{black}{Nonlinear indices n$_2$ are found to be in the range from $10^{-17}$ m$^2$/W to $4\times10^{-15}$ m$^2$/W, for $n=3$ to $n=7$.
%
%Response times found for nonlinearities at n = 4 are in the range below 40ps and system recovery time is in the nanosecond range.
In single pulse experiment the response time of nonlinearity must be given by the polariton lifetime comparable to the duration of the pulse $\sim1$ ps. Additional pump-probe measurements reveal that the nonlinearities at $n = 4$ are found to respond within a picosecond rise time and fall within $\sim40$~ps range. This ultrafast response is followed by additional nonlinear dynamics rising and falling on density-dependent timescales of order $100$~ps to $2$~ns. This demonstrates that multiple processes contribute to the nonlinear response.
} 
Crucially, our pulsed resonant excitation method \textcolor{black}{significantly reduces the interactions with long-lived ground state excitations and electron-hole plasma and enables us to access the pure, ultra-fast, Rydberg nonlinearity.
%
%that can provide otherwise inaccurate measurement of pure Rydberg nonlinearity.
To complete our study we provide a theoretical model that takes into account contributions from Rydberg and Pauli blockade,} and semi-quantitatively explains the observed experimental behaviour at high excitation densities.

%----
\section*{Single pulse experiment}

\textit{Experiment.} We study a microcavity system formed by two silver mirrors with an embedded thin flake of \cutwoo material deposited on top of SiO$_{2}$ substrate with an intermediate PMMA layer (see Fig.~\ref{fig:overview}\textbf{a,b}). Multiple cavity modes with a free spectral range of $9$~meV form (see SI, Fig.~S6) due to the thickness of \cutwoo slab of approximately $26~\mu$m.
We reveal the strong coupling between the cavity modes and Rydberg excitons in the angular-resolved transmission spectra of a super-continuum laser source (see Fig.~\ref{fig:overview}\textbf{c}), similar to recently published results~\cite{Orfanakis2022}. The transmission spectra recorded for different in-plane $\bm{k}$-vectors show anti-crossings between the cavity modes and Rydberg exciton resonances in Fig.~\ref{fig:overview}\textbf{c} with clear doublets corresponding to the upper (UP) and lower (LP) exciton-polariton states. To highlight the strong coupling in Fig.~\ref{fig:overview}\textbf{d} we also plot the dispersion of Rydberg exciton-polaritons for $n = 3$, $4$ and $5$. The formation of polaritons for $n = 6$ and $7$ excitons is not resolved for this position on the sample, since the cavity mode is in resonance at higher wavevectors and the high frequency spectral noise leads to weak signal to noise ratio and in $\bm{k}$-space prevents observation of the anticrossing. The signal to noise ratio is higher for detection in real space (i.e Fourier transform of $\bm{k}$-space) and the polariton doublets for $n = 6$ and $7$ are clearly observed in transmission for a different spot on the sample with a slightly different {\cutwoo} thickness, when the energy of the cavity mode at $k=0$ is tuned into resonance with these excitons by changing position on the sample (SI, Sec.~4).

To probe Rydberg exciton-polariton nonlinearities we record the transmission spectra in real space at different energies of the laser pulse (with full width at half maxima FWHM~1.75 meV), directed at normal incidence to the sample and tuned in resonance with the polariton states arising from excitons with different $n$. 
In Fig.~\ref{fig:waterfall_spectra} we show the power dependencies of the transmission spectra of these states. At small powers the doublet of LP and UP states is clearly visible for quantum states from $n=3$ to $n=7$.
\textcolor{black}{There is a minor asymmetry between the intensities of the lower and upper polariton branches observed for n=4 and 6 states, which is attributed to a small detuning between the laser peak energy and the centre between polariton resonances (this detuning is set manually by the diffraction grating in the pulse shaper and cannot be very precise), so that one resonance is pumped slightly more efficiently than the other.}

As power increases the separation between the polariton resonances becomes smaller, which we attribute to the decrease in coupling strength, and eventually a collapse of strong coupling. The resulting transmission spectra profiles become a singlet (here, limited by the pulse spectral width). We note the sharp contrast in power threshold needed to reach a singlet between the $n = 3$ case (of the order $100$~nW or 500~$\upmu$J/cm$^2$) and the $n = 7$ case (of the order of $1$~nW or 5~$\upmu$J/cm$^2$). \textcolor{black}{We exclude excitation induced thermal effects as no red shifts of the exciton resonances were detected in the experiments (for details see SI, Sec. 3).}

The fitting of the spectra at each power is performed using a coupled oscillators model previously used for Rydberg exciton-polaritons \cite{Orfanakis2022}, taking into account the spectral profile of the pulse and with the coupling strength being the main fitting parameter (see Methods and SI, Sec. 1).
The model fits the experimental data well, in particular at low powers. 
From these fits we can extract the Rabi splitting $\Omega$ as a function of exciton density $\rho$, which is plotted in Fig.~\ref{fig:coupling_strengths}  (\textcolor{black}{see Methods for equation to deduce $\rho$ and SI, Sec.~5 for its derivation)}. {\color{black}We stress that $\rho$ being the excitonic fraction of resonantly pumped polaritons is an important parameter that defines the absolute value of nonlinear energy shifts}. At low pump power the overall Rabi splitting drops with \textcolor{black}{quantum number} $n$, since larger exciton size leads to smaller oscillator strength \cite{HeckotterPRB2017}.
%, as electron-hole complexes become delocalised. 
The Rabi splitting is also observed to decrease strongly with $\rho$ or pumping power, showing a fast initial drop and overall nonlinear scaling with density at larger occupations. {\color{black}We also note that deducing this drop becomes difficult at larger $n$ as the absolute value of the linear light-matter coupling decreases, while cavity decay rate remains the same.}

%{\color{black} 
%}
 
 %\textcolor{black}{The exciton-polariton Kerr-like nonlinearity due to quenching of Rabi splitting is usually characterised by coefficient $\beta=d\Omega/d\rho$ \cite{Emmanuele2020,Brichkin2011,Zhang2022}, which, in other words, determines the energy renormalisation of the polariton branches with density. In this sense, $\beta$ is directly related to the nonlinear refractive index coefficient of the cavity active region (\cutwoo), since change of the cavity refractive index leads to the change of the polariton mode energies ( see Ref.\cite{Emmanuele2020} and references therein and SI.).  }

\textcolor{black}{The exciton-polariton Kerr-like nonlinearity due to reduction of Rabi splitting $\Omega$ with density $\rho$ is usually characterised by the coefficient $\beta=d\Omega/d\rho$ \cite{Emmanuele2020,Brichkin2011,Zhang2022}. Classically, the Rabi splitting arises because the excitonic oscillators in the \cutwoo active region add a frequency-dependent contribution to the refractive index which modifies the cavity resonance condition~\cite{Khitrova:RMP71(1999)}. The magnitude of this refractive index component, as well as the Rabi splitting, decreases as $\rho$ grows. Thus we can directly relate the nonlinear refractive index of \cutwoo to $\beta$ (see Ref.~\cite{Emmanuele2020,Walker2015} and references therein, and SI).}
 
 Here we focus on the nonlinear response at lower densities, when the polariton doublets are still resolved and $\Omega$ behaves nearly linearly with $\rho$. 
The deduced $\beta$ factors are plotted in Fig.~\ref{fig:coupling_strengths}\textbf{b} as a function of principal quantum number $n$ of polaritonic states (black dots, labelled as $\beta_{\mathrm{exp}}$). We observe rapid increase of nonlinearity as a function of $n$ (note the log-linear scale), and the fitting provides scaling $\beta_{\mathrm{exp}}\sim n^{4.4\pm 1.8}$ (red solid curve in Fig.~\ref{fig:coupling_strengths}\textbf{b}) (see SI, Sec.~7.1.). The $\beta$-values are found to be between 0.01~$\upmu$eV$\upmu$m$^3$ for $n=3$, to 0.4~$\upmu$eV$\upmu$m$^3$ for $n=7$ (note, here we use volume units as natural for bulk crystal in a cavity). Already for $n=5$ this nonlinearity exceeds that in other highly interacting polariton platforms, such as microcavities with GaAs-based quantum wells, if one takes into account excitonic spatial confinement (see SI, Sec.~6). 
Below we use the experimentally obtained $\beta$ values and scaling to analyse main potential contributions and compare with our theoretical model.
%%%

%%%

\textit{Theoretical analysis.} {\color{black}To explain the experimental results we develop a theory for describing the effective decrease of light-matter coupling. Specifically, we take into account dipole-dipole interactions between Rydberg states of p-wave excitons, which are known to reduce \cutwoo absorption  with increasing exciton density in the cavity-free case \cite{Amann2020}. This phenomenon may be explained as the formation of a blockade region in the vicinity of spatially extended exciton. In this region, light can no longer create new excitons due to the strong dipole-dipole interaction shifting the exciton energy out-of-resonance. Consequently, the optically active region in a sample decreases with the increase of exciton density.}

{\color{black}We model this blockade effect and }plot the $\beta_{\mathrm{Rydberg}}$ dependence using the theoretically predicted values of dipole-dipole interactions~\cite{Walther:PRB98(2018)} (solid purple curve in Fig.~\ref{fig:coupling_strengths}\textbf{b}). We observe that $\beta_{\mathrm{Rydberg}}$ scales as $\sim n^{5.5}$ (SI, Sec.~7.2). {\color{black}In this theoretical plot, a full blockade is assumed where the blockade region is defined sharply by a step-like boundary at the Rydberg blockade radius ($r_C$). In reality, the blockade effect is coming from the exciton’s density-density correlations such that transition between blockaded and non-blockaded regions is smooth~\cite{Hecktter2021}, meaning that the creation of additional exciton within $r_C$ is not strictly forbidden. Therefore, the full blockade result gives approximately the upper bound which is likely to overestimate the nonlinearity. Furthermore,} since the dipole-dipole interaction constants are generally difficult to calculate exactly and may be also overestimated \cite{Walther:PRB98(2018),Amann2020}. To compare, we plot a scaled line (by a factor of 1/5) for the strength of dipole-dipole interaction, which matches the overall trend for experimental nonlinearity and provides a good fit (dotted purple curve in Fig.~\ref{fig:coupling_strengths}\textbf{b}). 
%%%

Another possible origin of the reduction of Rabi splitting with increasing density is a nonlinear phase-space filling (NPSF) in polaritonic systems, as commonly observed with ground state s-excitons~\cite{Emmanuele2020,Zhang2022}. {\color{black} Nonlinear phase-space filling, also known as nonlinear saturation, is the statistical effect that emerges from the non-bosonic behavior of excitons at large occupations. As excitons are composite quasiparticles, the Pauli blockade prevents excitation of certain excitonic configurations if they are already occupied. This nonlinear decrease in the density of states leads to the effective reduction of light-matter coupling. Similarly to the Rydberg-induced case, the nonlinearity also grows with the exciton size, as a smaller number of excitons can be created (per volume or area) until the medium becomes effectively transparent.} The blue curve in Fig.~\ref{fig:coupling_strengths}\textbf{b} shows the scaling of Pauli-induced nonlinearity for the experimentally deduced Bohr radius $a_{\mathrm{0}} = 0.83$~nm (for $n=1$, see \textcolor{black}{SI, Sec.~7.2}).
The $\beta$ factor associated to the Pauli blockade ($\beta_{\mathrm{Pauli}}$) has an asymptotic $\sim n^{2.5}$ scaling, while at low $n$ it is described by a $\sim n^{3.5}$ dependence (SI, Sec.~7.2).
The Pauli blockade curve, which has no fitting parameters, is well below the experimental values. To at least partially fit the experimental values using only the Pauli blockade, the Bohr radius must be set to $a_{\mathrm{0}} \approx 2$~nm. This greatly exceeds the exciton radius estimates from the measured low-density absorption (SI, Sec.~7). Therefore, we conclude that the Pauli blockade alone cannot explain the observed nonlinearity and that the Rydberg-induced blockade plays a dominant role. The contribution of the Rydberg blockade to exciton-polariton nonlinearity is an order of magnitude stronger (see Fig.~\ref{fig:coupling_strengths}\textbf{b}).

\section*{Nonlinear n$_2$ parameter}

\textcolor{black}{From the nonlinear optics perspective the polariton nonlinearity can be also characterised by the nonlinear refractive index n$_2$ of the active medium in a microcavity (Cu$_2$O in our case)~\cite{Emmanuele2020}. This nonlinear parameter appears in the total refractive index as a frequency- and intensity-dependent term, $\mathrm{n_T}(\omega) = \mathrm{n}_0(\omega) + \mathrm{n}_2(\omega) I$. The n$_2$ parameter from blockade effects may be estimated by using Eq.~\eqref{eqn:n2} (see SI, Sec~7.3):
\begin{equation}\label{eqn:n2}
    \mathrm{n}_2(\omega)\approx -\frac{ h_n  \beta_n }{2 c \mathrm{n}_0^2 \omega }\frac{G_n^{(0)}(\omega-\omega_n)}{(\omega-\omega_n)^2+\frac{1}{4}\gamma_n^2},
\end{equation}
where $G_n^{(0)}$ is the light matter coupling constant, $\omega_n$ is the Rydberg resonance frequency, $\gamma_n$ is the excitonic linewidth, $n_0$ is the background refractive index of Cu$_2$O, $c$ is the speed of light in vacuum and $h_n$ is a constant of proportionality (determined from the measured Rabi splitting using Eq.~(S25), see SI, Sec.~7.3). Using the $\beta$-factors in Fig.~\ref{fig:coupling_strengths}b, we derive the energy dependencies of n$_2$ for each individual excitonic mode in Fig.~\ref{fig:n2}. Black dots show the upper bounds of n$_2$ obtained using the theoretical estimates of $\beta$, with values ranging from $10^{-17}$ to $10^{-15}$ m$^2$/W, for $n=3$ to $n=7$. % which agree with estimations by using the method in Ref.\cite{Emmanuele2020} (Fig. \ref{fig:n2_param}).
Using the experimentally measured $\beta$-factors we deduce peak n$_2$ values ranging from $10^{-17}$ m$^2$/W to $4\times10^{-15}$ m$^2$/W, for $n=3$ to $n=7$. The peak values measured for $n=7$ exciton resonance are comparable to n$_2$ in GaAs polariton waveguides \cite{Walker2015}.
} 
 
\textcolor{black}{These values of n$_2$ are 8 orders of magnitudes lower than those measured on bulk Cu$_2$O in Ref.~\cite{Morin2022}, where CW excitation was used in resonance with Rydberg excitons. However, we note that studies of polariton nonlinearity in GaAs-based photonic systems showed that CW excitation usually results in observed effective n$_2$ values one-two orders of magnitude bigger than in the case of picosecond pulsed pumping. Such a discrepancy was explained by population of long-lived (up to $100$~ps) excitons by the CW beam, the interaction with which leads to enhanced nonlinear energy shift \cite{PWalker2017}.}
\textcolor{black}{
In the case of \cutwoo the CW excitation in the vicinity of Rydberg exciton resonances and the subsequent relaxation of the photoexcited carriers to the ground 1s level is expected to lead to high population of long-lived (with a lifetime of $13~\upmu$s \cite{MysyPRL1979}) dark 1s paraexcitons. Lifetimes up to several hundreds microseconds were also reported, which was attributed to long-lived excitons trapped at defects or unknown metallic impurities \cite{Rogers2022}. 
At sufficiently high density long lived 1s excitons or excitons trapped to defects may further recombine through Auger recombination creating plasma. The exact density of 1s excitons and plasma in this case may depend on the particular sample and the number of defects at which 1s excitons may accumulate.} 

\textcolor{black}{A possible explanation for the very high values of n$_2$ observed in Ref.~\cite{Morin2022} is therefore an interplay between resonantly pumped Rydberg excitons and plasma. Free carriers may increase the strength of Rydberg blockade through screening, which increases the size of the Rydberg excitons leading to enhancement of the dipole-dipole interactions \cite{WaltherPRL2020}. At the same time, the screening may reduce the oscillator strength of the excitons resulting in a density dependent change in their contribution to the refractive index.
%screening of the Rydberg excitons may lead to the increase of its size and hence may enhance the dipole-dipole interactions (i.e Rydberg blockade regime) \cite{WaltherPRL2020}.
In support of this suggestion we note that Heck\"{o}tter et. al. \cite{HeckotterPRL2018} characterised the influence of free carriers, showing that absorption for $n=10$ Rydberg excitons reduces by a factor of $\sim3$ at free carrier density as low as $\sim 0.5$ $\upmu$m$^{-3}$.
}
\textcolor{black}{
In the following section we use time-resolved measurements to shed some light on the regimes where the polariton nonlinearity is dominated by ultra-fast processes, such as Rydberg blockade, and where it may be complicated by slower processes such as Auger mediated generation of plasma.
}
%Finally, it is also important to note that polaritons are distinct from and more complicated than excitons, being a superposition state relying on coherent energy exchange between excitonic dipoles and electromagnetic waves. It is therefore perhaps not surprising that they could exhibit different nonlinear properties.

\section*{Pump-probe experiment}

\textcolor{black}{
Importantly, in the single pulse experiment presented above the polariton nonlinearity that we measured must be ultrafast since it develops within the picosecond timescales of the probe pulse (which is also the pump pulse in that case) and the polariton lifetime. However, some of the resonantly excited exciton-polaritons can be absorbed, with the resultant formation of lower energy excitons, and then plasma through the Auger exciton recombination. Therefore the Rabi splitting may remain quenched for some time after the pump pulse is gone due to interaction with this long-lived plasma. 
To reveal this effect we further perform pump-probe measurement of Rabi splitting for the $n=4$ resonance. The $n=4$ polaritons were excited with a strong pump and the transmission of the sample was measured using much weaker probe pulse delayed by some time from the pump. The linear polarisation of the probe is chosen to be perpendicular to the pump and the transmitted pump beam was rejected by a linear polariser. The $n=4$ state was chosen because it provides a good signal to noise ratio for small probe powers.}

\textcolor{black}{
We plot the results of the pump-probe experiment in Fig.~\ref{fig:pumpprobe}. In Figs.~\ref{fig:pumpprobe}a and \ref{fig:pumpprobe}b we show the transmission spectra for selected delays between pump and probe pulses for pump fluences of 100~$\upmu$J cm$^{-2}$ and 450 $\upmu$J cm$^{-2}$, respectively. The probe fluence was 50~$\upmu$J cm$^{-2}$. The transmission spectra are then fitted to extract the Rabi splittings as a function of the time delay. In this fitting procedure we omit those time delays between $-29$ and $37$ ps (apart from zero delay), for which the interference between the residual pump and probe prevents reliable fitting (the residual pump intensity in the polarisation of the probe is about 30 times less than that of the probe for pump fluence 450 $\upmu$J cm$^{-2}$ and can not be suppressed completely by the linear polariser, see SI, Sec.~9).}

\textcolor{black}{At pump fluence of 100~$\upmu$J cm$^{-2}$ the Rabi splitting is reduced at $t=0$ as expected for the instantaneous Rydberg exciton-polariton blockade mechanism and by $t \approx 40$ ps has recovered almost back to its value prior to the pump arrival (Fig.~\ref{fig:pumpprobe}c, bottom panel). At longer delay times $t>40$ ps the Rabi splitting reduces and then increases again on a time scale of $200$-$300$~ps. We attribute this behaviour to the 1s exciton and plasma dynamics created by the pump pulse, which is expected to occur on a nanosecond timescale at exciton densities $10^3$-$10^4$ $\upmu$m$^{-3}$ 
given the Auger recombination rate of 1s excitons is of the order of $10^{-4}$ $\upmu$m$^{3}$/ns \cite{StolzPRB2022}. At higher pump fluences of 450 $\upmu$J cm$^{-2}$ and 1000 $\upmu$J cm$^{-2}$, shown in the middle and upper panels of Fig.~\ref{fig:pumpprobe}c, the creation of plasma after the pump pulse is expected to occur on a shorter timescale and indeed we observe the Rabi splitting is rapidly quenched and then recovers monotonically on timescales of $1.2$ and $1.8$ ns, respectively. The exact explanation of the observed temporal dynamics requires complex modelling of the exciton-plasma conversion and decay using the corresponding rate equations \cite{StolzPRB2022} and is beyond the scope of this manuscript. } 

\textcolor{black}{Finally, we also study Rydberg exciton-polariton transmission spectra under non-resonant CW excitation well above the band gap of the Cu$_2$O. In this case, we find that the collapse of the Rabi-splitting is observed at photon densities about 6 orders of magnitude less than in the pulsed excitation regime (see SI, Sec. 8). This experiment further confirms the important role of long-lived 1s states and plasma, the density of which can be significantly higher in CW than pulsed excitation.}

%high powers($< 40$ ps, the exact values for the rise time cannot be extracted due to residual interference between pump and probe pulses) and the recovery time $\tau_R$ is increasing with power up to 1.8 ns. 

%These times are much shorted than the cycling time of 1 ms and can be explained by population of 1s states and further plasma creation trough Auger process that affects nonlinear response. Clearly in our our case of pulse pumping of polariton states with a laser of $1$~kHz repetition rate we probe the Kerr-like non-linearity due to instantaneous interactions between Rydberg exciton-polaritons within the short time duration and the system relaxes to the ground state after each pulse which is not the case in CW pumping experiments

\section*{Conclusion} 
\textcolor{black}{In conclusion, for the first time we investigated the nonlinear behaviour of a polaritonic system based on Cu$_2$O in the ultra-fast pulsed regime. The polariton system allows us to probe} the collapse of the Rabi splitting for Rydberg exciton-polaritons and we observed that the associated polariton nonlinearity increases as $\sim n^{4.4\pm 1.8}$ with principal quantum number $n$ due to stronger dipolar exciton-exciton interactions. The experimental values of polariton nonlinearity coefficient $\beta$ are in good agreement (within a factor of $5$) with our microscopical model, which takes into account both the Rydberg dipole-dipole interactions and Pauli blockade without fitting parameters. 
\textcolor{black}{Furthermore, our pump-probe data suggests that there are several contributions to the nonlinearity in Cu$_2$O based systems, which act on several different timescales. As well as the ultra-fast response there are contributions which can persist significantly longer than pulse duration and polariton lifetime. The timescales for these are consistent with population of long lived states and creation of plasma. In the CW excitation regime these can be even more pronounced resulting in greatly enhanced nonlinearity. In order to investigate the effect of plasma on Rydberg exciton or exciton-polariton blockade in more detail, and to fully explain the drastic differences observed in nonlinearities for pulsed and CW excitation, we suggest further studies using for example pulsed excitation with varying pulse duration and repetition rate.
}

We note that higher exciton-polariton nonlinearities are possible to achieve by modification of several factors: (i) use higher quality crystals and observe higher $n$ states, (ii) use higher quality microcavities to increase strong coupling with high $n$ states, (iii) reduce the thickness of \cutwoo to quantum well level to decrease absorption losses in the cavity and enhance exciton-exciton interactions, and (iv) exploit electromagnetically induced transparency for the reduction of losses due to phonons~\cite{Walther2018}.
Our work demonstrates that Rydberg polaritons in Cu$_2$O is a suitable platform for quantum polaritonics with nonlinearities that scale sufficiently strongly with Rydberg exciton quantum number to reach the single polariton nonlinearity.

%-------------------

%TC:break methods
\section*{Methods}\label{sec5}
\textit{Sample and setup.} Our cavity containing a natural {\cutwoo} sample is cooled down to 4 K in a continuous flow liquid Helium cryostat. Natural {\cutwoo} crystals employed here since these are of higher quality than artificially grown samples \cite{Kazimierczuk2014,PhysRevMaterials.5.084602}.
To prepare the microcavity sample, the Cu$_2$O flake was first cleaned in xylene with 1 minute of delicate sonication and rinsed in isopropanol. Producing the mirror layers was done by initially securing the Cu$_2$O flakes onto a substrate with a 45 mg/ml solution of PMMA and toluene. The sample with the Cu$_2$O flake was then loaded into an Ångstrom Engineering thermal evaporator. Silver was then evaporated onto the Cu$_2$O using a resistive source at a rate of 0.2 nm $s^{-1}$. The final thickness of the silver mirror was 50 nm. The sample was then unloaded and the Cu$_2$O flake was removed and carefully rotated to expose the opposite side that has no silver deposited. The PMMA solution was then used to secure this to a new substrate and the deposition was repeated with the same thickness. To attach the resulting structure on the final substrate, PMMA 495 resist in 8 \% anisole was spun at 4000 rpm on a glass slide resulting in a 600 nm thick layer. The flake was transferred onto the resist, and then baked for 5 min at 180 $^{\circ}$C.

\begin{figure*}
\centering
\includegraphics[width=1\linewidth]{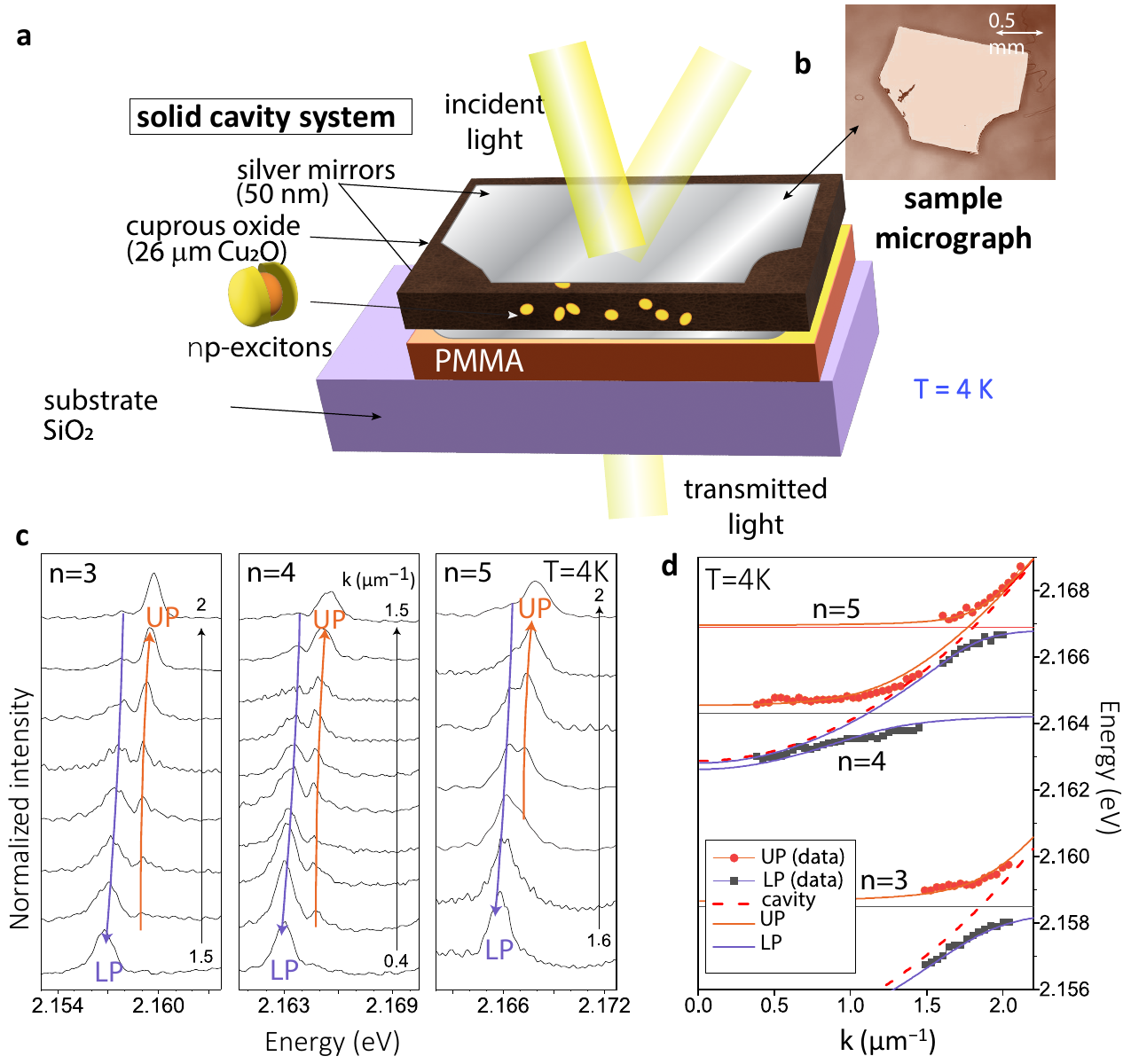}
    \caption{\textbf{Sketch of the {\cutwoo} microcavity sample and its dispersion.} \textbf{a} A schematic of solid cavity used for the transmission experiment. \textbf{b} Microscopy image of the natural {\cutwoo} sample of thickness approximately $26~\mu$m with silver mirrors (recoloured). \textbf{c} Waterfall spectra plots taken with super-continuum laser in transmission for different wavevectors near $n=3$, $n=4$ and $n=5$ excitons. Upper and lower polaritons are visible as a doublet at resonant conditions. \textbf{d}  Polariton peak positions as a function of wavevector extracted from \textbf{c}. Lines show cavity, exciton and polariton dispersions (for higher principal quantum numbers, see SI, Sec. 4.) }
    \label{fig:overview}
\end{figure*}

\begin{figure*}[th]
    \centering
    \includegraphics[width=1\linewidth]{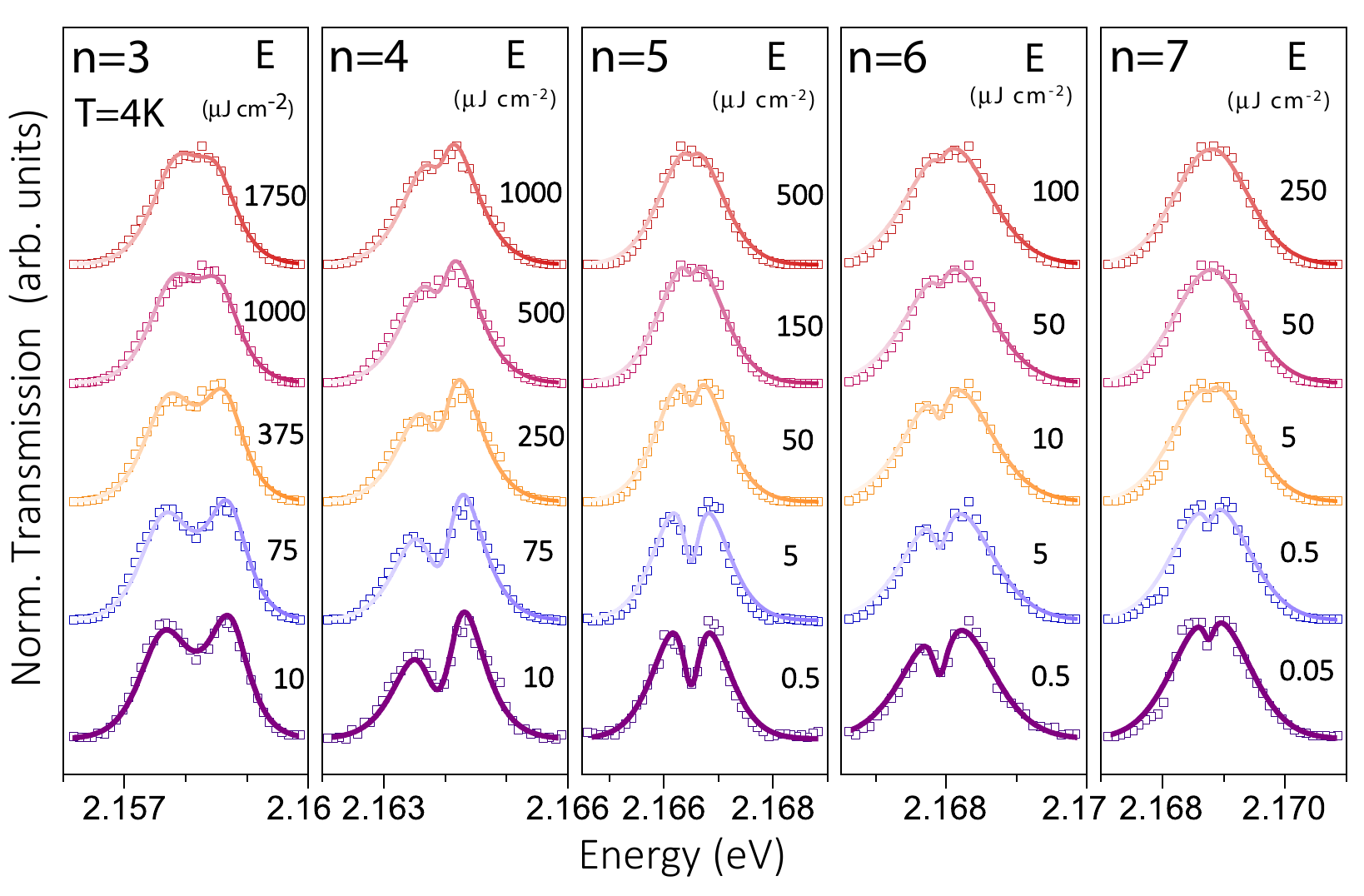}
    \caption{\textbf{Normalised transmission spectra with increasing pump power.} We plot the transmission of the {\cutwoo} microcavity from narrowband ($1.75$~meV) excitation (see Methods) for different principal quantum numbers $n$, taken at different \textcolor{black}{ pump fluence $E=P_{\mathrm{avg}}/(f A)$ ($P_{avg}$ is pump incident average power, $f$ is the repetition rate of the laser, $A$ is the effective excitation area). } The spectra are normalised to unity and shifted by 1 for clarity. The solid curves represent a fit with a coupled oscillators model.}
    \label{fig:waterfall_spectra}
\end{figure*}

\begin{figure}[ht!]
    \centering
    \includegraphics[width=0.7\linewidth]{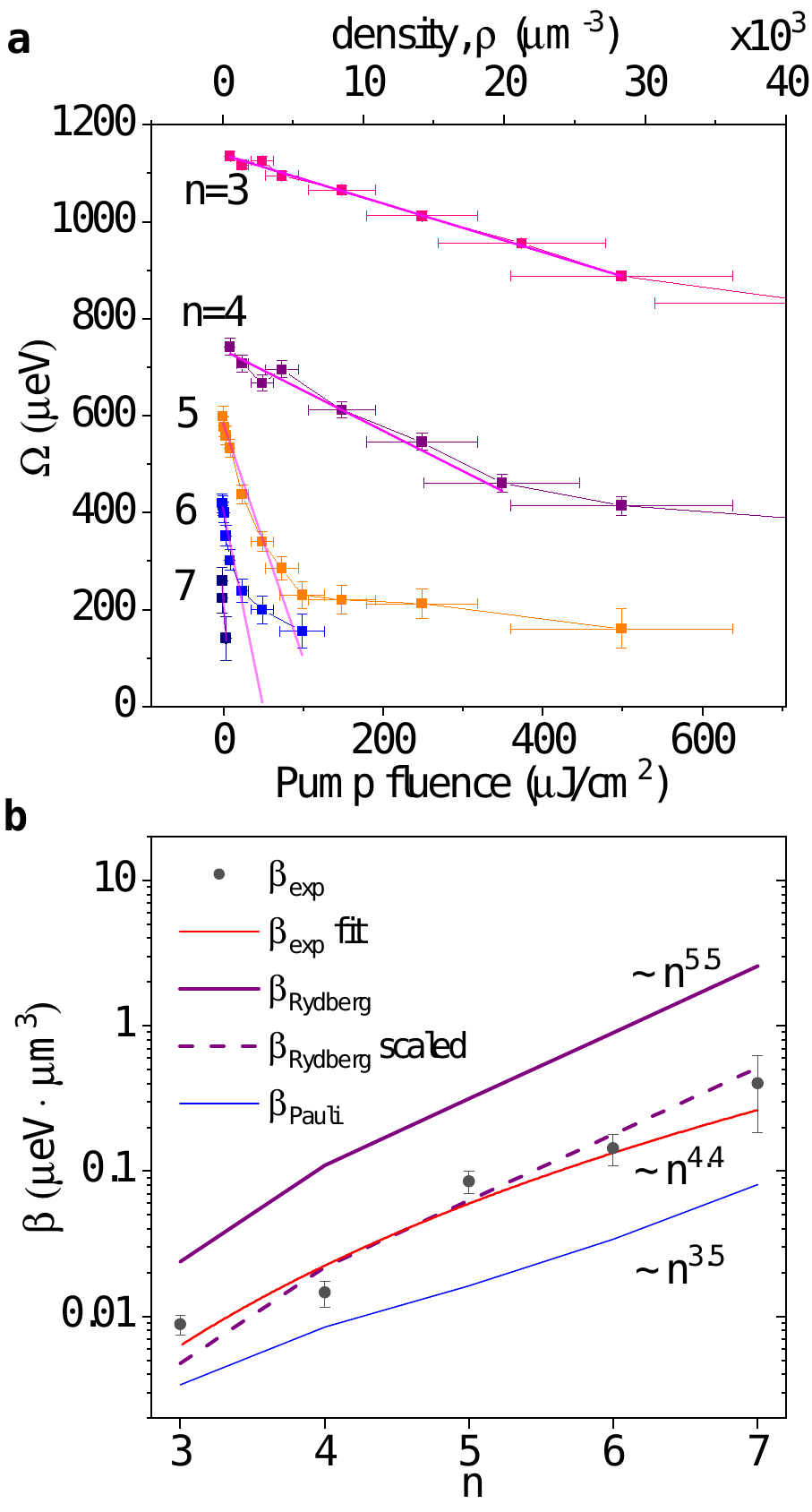}
    \caption{\textbf{Nonlinear Rabi splitting decrease and its scaling. a} Rabi splitting vs exciton density $\rho$ \textcolor{black}{and/or pump fluence} is shown for different principal quantum numbers $n$, and reveals an $n$-dependent nonlinear coefficient. Data are extracted from spectra in Fig.~\ref{fig:waterfall_spectra}. The lines show the linear trend seen at low density. \textbf{b} Linear-log plot of nonlinear coefficients ($\beta$-factors) vs principal quantum number $n$. Filled dots correspond to the experimental $\beta$-factor extracted from Fig.~\ref{fig:coupling_strengths}\textbf{a}, which can be fitted with an $ n^{4.4 \pm 1.8}$ dependence (solid red curve, error for 90\% confidence level). The dash purple and solid blue curves show the calculated $\beta$-factors for Rydberg and Pauli blockade, respectively. The solid purple curve depicts Rydberg saturation scaling for the full blockade ($C_6 \approx 10^{-15}~n^{11}$~meV$\mu$m$^6$). (All calculated error bars are for standard error.)
    }
    \label{fig:coupling_strengths}
\end{figure}

\begin{figure}[ht!]
    \centering
    \includegraphics[width=0.8\linewidth]{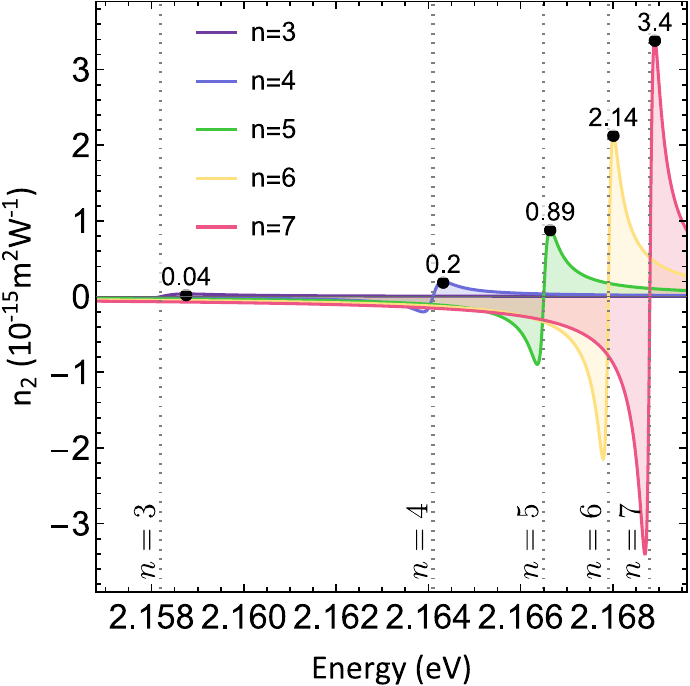}
    \caption{\textcolor{black}{\textbf{Nonlinear refractive index n$_2$.} Coloured curves show contributions from each excitonic mode with $n=3,\dots,7$. Here we use $\beta = 1.41\times10^{-5} n^{5.5} \upmu \mathrm{eV}\mu \mathrm{m}^3$ corresponding to the dashed purple curve in Fig.~\ref{fig:coupling_strengths}.
    %which is obtained by using the data in Fig.~\ref{fig:coupling_strengths}}
    }}
    \label{fig:n2}
\end{figure}

\begin{figure*}[ht!]
    \centering
    \includegraphics[width=1\linewidth]{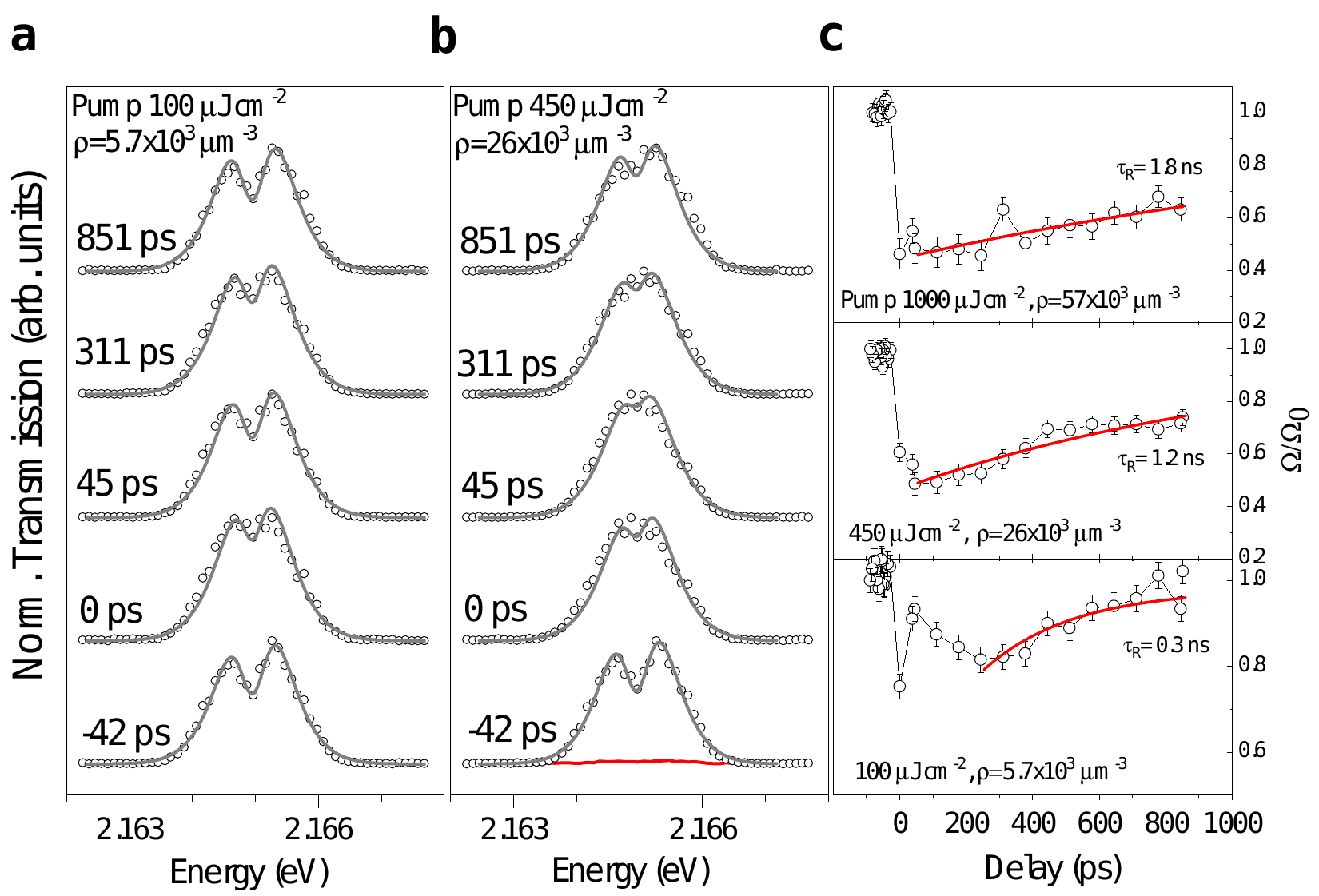}
    \caption{\textcolor{black}{\textbf{Pump-probe of the Rabi splitting at $n=4$. a} Normalised transmission spectra at different delay times between pump and probe pulses at fixed pump(probe) fluence 100~$\upmu$J cm$^{-2}$ (50~$\upmu$J cm$^{-2}$). Solid grey lines correspond to fits. \textbf{b} Normalised transmission spectra at different delay times between pump and probe pulses at fixed pump(probe) fluence 450~$\upmu$J cm$^{-2}$ (50~$\upmu$J cm$^{-2}$). Red line corresponds to pump only transmission signal ($\approx 30$ times smaller than the probe after polarisation rejection). Solid grey lines correspond to fits. \textbf{c} Rabi splitting normalised by Rabi splitting at negative time delays ($\Omega/\Omega_{0}$) extracted from the spectra and plotted as a function of delay time between pump and probe at three pump fluences (100, 450 and 1000 $\upmu$J cm$^{-2}$). Solid red lines are single exponent fits.}}
    \label{fig:pumpprobe}
\end{figure*}

The Fourier space imaging spectra in Fig.~\ref{fig:overview} have been obtained with a 1 ns super-continuum laser with a repetition rate of 23 kHz and a spectral width after filtering of 50 nm. In order to achieve narrow band resonant excitation in Fig.~\ref{fig:waterfall_spectra}, we used 100 fs pulses at 1 kHz repetition rate obtained from the frequency doubled output of a TOPAS optical parametric amplifer and then filtered by a 4f configuration pulse shaper with a 1200 g/mm grating and a slit slightly displaced from the focal plane to obtain a Gaussian shape spectrum with FWHM~1.75 meV. 

\textit{Fitting procedure and extraction of $\beta$ factors.} 
The transmission spectra resulting from resonant excitation consists of a doublet centered around the excitonic resonance. In order to extract the coupling strength of the Rydberg polariton from the transmission spectra, we fit the data according to the model used in \cite{Orfanakis2022} taking also into account the small spectral width (FWHM=1.75 meV) of the excitation pulse [see SI, Sec. 1, Eq. (S1)].
Cavity and excitons linewidths in the fit were fixed for all powers and obtained in a separate measurements for each $n$ (see SI, Sec. 2). Two fitting parameters were allowed to vary with power: the amplitude and the coupling strength. Thus we extract the coupling strength dependence on average power.
To extract $\beta$ we plot the Rabi splitting as a function of resonantly excited exciton density estimated from the transmitted power through the cavity in resonant conditions and fit it with the linear function. The slope provides an average estimate for $\beta$. 

The density of excitons $\rho$ per volume is calculated from the incident average excitation power $P_{\mathrm{avg}}$ by using the following equation:
\begin{equation}
    \rho =\frac{T P_{\mathrm{avg}}\tau'\lvert X\rvert^{2}}{f \tau_{\mathrm{p}} L A \hbar \omega \lvert C\rvert^{2}},
\end{equation}
where $T=1/180$ is the fraction of incident power that is transmitted, $\tau'=14$ ps is the inverse of the tunnelling rate of photons out of the cavity, through the silver mirror and towards the detector, $f= 1$~kHz is the laser repetition rate, $\tau_{\mathrm{p}} \approx 1.57$~ps is the effective laser pulse width, $A\approx20$ $\mu$m$^2$ is the effective area for the interaction \cite{Kuriakose2022}, $L = 26~\mu$m is the cavity length, $\hbar\omega$ is the single photon energy and $\lvert C\rvert^{2}=0.5$ and $\lvert X\rvert^{2}=0.5$ are photonic and excitonic fractions of the polaritons respectively (see SI, Sec. 5 for more details).

%TC:ignore
\textbf{Acknowledgments}

We acknowledge UK EPSRC grants EP/V026496/1, EP/S014403/1 and EP/S030751/1. OK and KWS acknowledge UK EPSRC grants EP/V00171X/1 and EP/X017222/1, and NATO SPS project MYP.G5860. HO acknowledges The Leverhulme Trust (Agreement No. RPG-2022-188).
%TC:endignore

%TC:break supplementary

\textbf{Supplementary information}

Supplementary data supporting the finding of this article is available below.

%TC:ignore
\textbf{Competing interests}

The authors declare no competing interest.

\textbf{Data availability}
The data that support the findings of this study are openly available from the University of Sheffield repository at https://doi.org/10.15131/shef.data.25225340 .% figshare?

\textbf{Code availability}
The code used for analysis is freely available from the University of Sheffield repository at https://doi.org/10.15131/shef.data.25233025 .

\textbf{Authors contributions}

MM, AD, KO, SKR designed and conducted the experiments. JH, TI, PC contributed to the sample fabrication. OK, KWS provided theoretical support. MM, AD, OK, KWS, PW, HO and DK wrote the manuscript. All authors contributed to the analysis and interpretation of the results and critical reading of the paper.


%%\appendix
%TC:endignore

%TC:

%%\nocite{*}
%\bibliography{sn-bibliography}

%\end{refsection}

%Word count
\newpage

%%%

%%%
\renewcommand{\thefigure}{\textbf{S\arabic{figure}}}
\renewcommand{\theequation}{S\arabic{equation}}

\newpage
\newpage
\clearpage

%\tableofcontents
\section*{SUPPLEMENTARY INFORMATION}

\localtableofcontents
\newpage
%TC:break maintext
\section{Transmission spectra fitting procedure}

To fit the transmission spectra we used the model of cavity transmission from Ref.~\cite{Orfanakis2022}, multiplied by a Gaussian function which accounts for the spectrum of the pulse incident on the cavity. This reads
\begin{equation}\label{eq:weighted_transmission}
    T_n\approx\frac{A\exp\left( -2 \left(\frac{\Delta_n}{\sigma}\right)^{\!2}\,\right)}{(\frac{\kappa}{2}+G_n^2\frac{\gamma_n/2+2Q_n\Delta_n}{\gamma_n^2/4+\Delta_n^2})^2+(\Delta_n-G_n^2\frac{\Delta_n-Q_n\gamma_n}{\gamma_n^2/4+\Delta_n^2})^2},
\end{equation}
where $A$ is the peak amplitude, $\Delta_n$ is the laser frequency detuning from the excitonic resonance with principal quantum number $n$, $\gamma_n$ is the excitonic linewidth for corresponding $n$, $\kappa$ is the cavity linewidth, $G_n$ is the coupling strength, $Q_n$ is the Fano asymmetry parameter, and $\sigma$ is the pulse spectral width. In all cases the subscript $n$ refers to the exciton with index $n$. We use Eq.~\eqref{eq:weighted_transmission} to fit the transmission spectra at each $n$ by fixing the parameters $\gamma_n$, $\kappa$ and $Q_n$ corresponding to each excitonic resonance taken from separate measurements (see Section 2 below). For each excitonic resonance we perform a global fit over all excitation powers, with only the coupling strength $G_n$ and the amplitude $A$ varying as a function of power. Thus we obtain power dependence of $G_n$.

\section{Measurement of excitonic and cavity linewidths}

\begin{figure}[H]
    \centering
    \includegraphics[width=0.5\columnwidth]{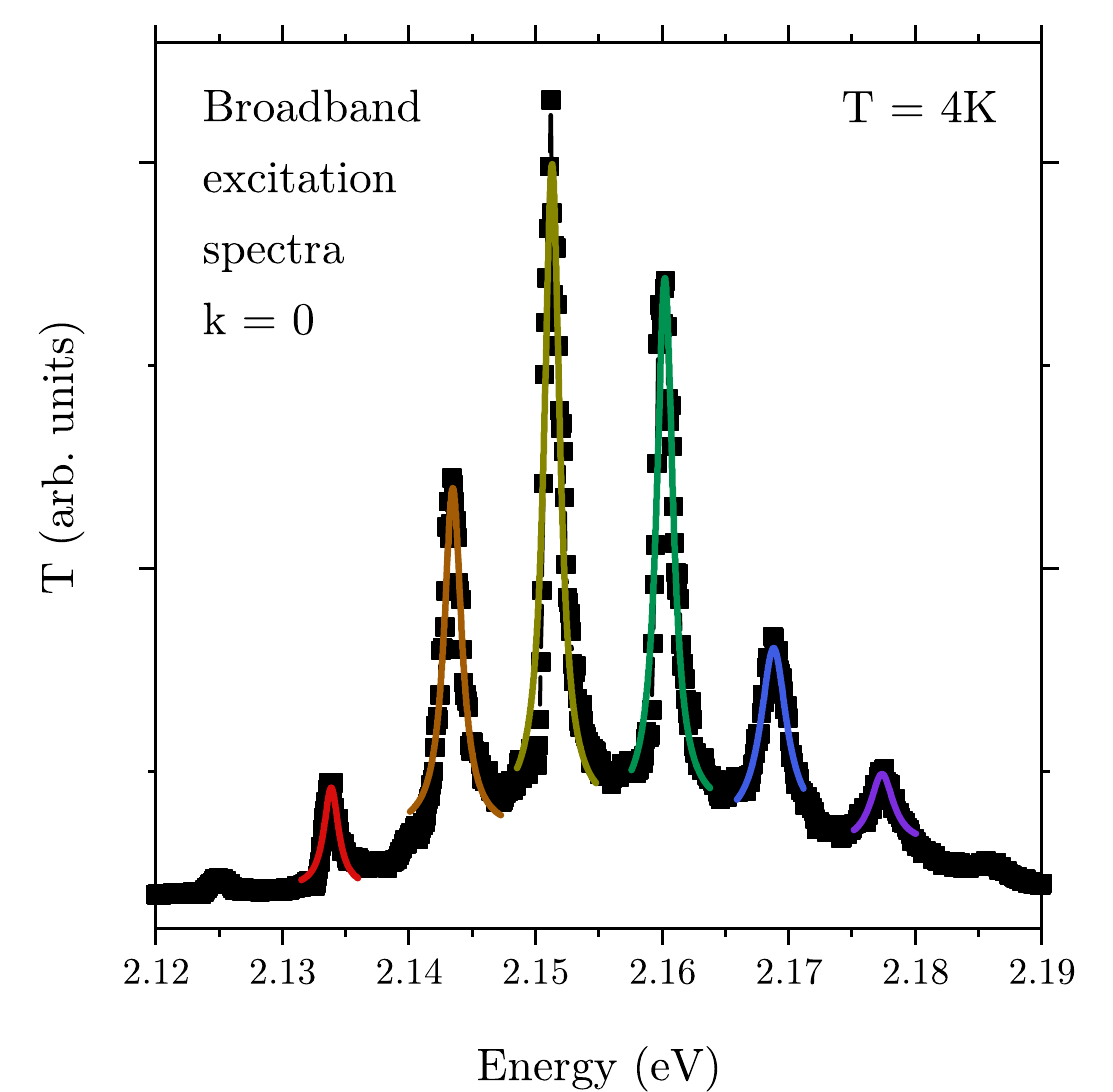}
    \caption{Normal incidence transmission spectrum of a {\cutwoo} microcavity under broadband excitation, showing uncoupled Fabry–Pérot cavity modes. Points are experimental data. Solid curves show Lorentzian lineshape fits to the modes.}
    \label{fig:cavity_fit}
\end{figure}

In Fig. \ref{fig:cavity_fit}, we extract the photonic cavity linewidth $\kappa$ of several modes at different energies. The data are taken from angle-resolved transmission (also called Fourier imaging or $\bm{k}$-space imaging) of a broadband super-continuum laser. A section of the angle-resolved data at zero incidence angle to the sample normal ($k=0$) is presented in the figure. As we want to extract an uncoupled and unperturbed ``purely photonic" cavity mode, the spectra are taken from a spatial region of the sample where the cavity modes are detuned far away from the excitonic resonances. This occurs due to a small wedge in the thickness of the sample.

\begin{figure}[H]
    \centering
    \includegraphics[width=0.55\columnwidth]{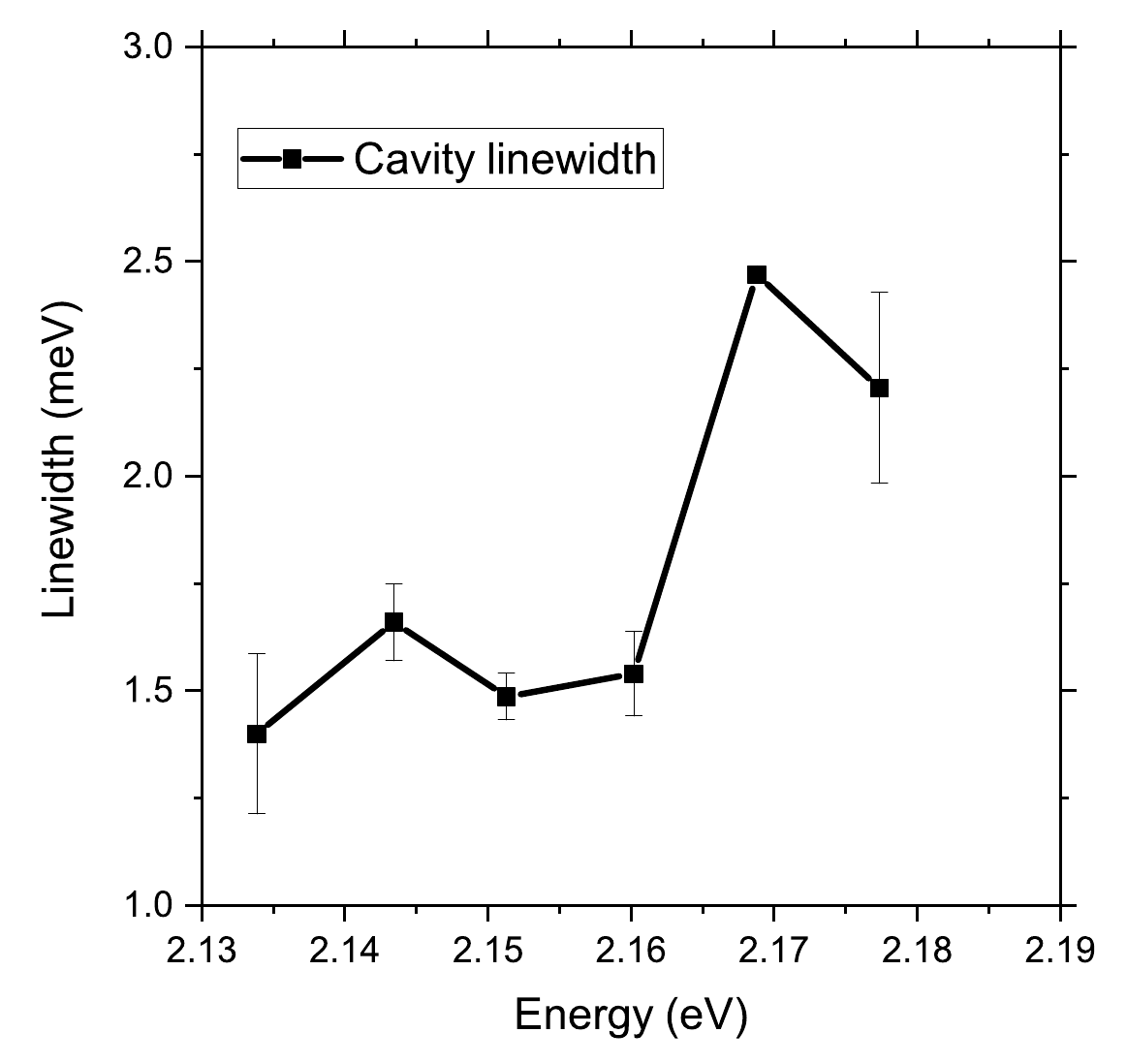}
    \caption{Cavity linewidth extracted from Fig. \ref{fig:cavity_fit} as a function of energy.}
    \label{fig:kappa_fit}
\end{figure}

From the Lorentzian fits in Fig. \ref{fig:cavity_fit}, we can extract the cavity linewidth as a function of energy, which we plot in Fig. \ref{fig:kappa_fit}. The trend can then be extrapolated to find appropriate $\kappa$ values for energy levels matching exciton resonances.

\begin{figure}[H]
    \centering
    \includegraphics[width=0.55\columnwidth]{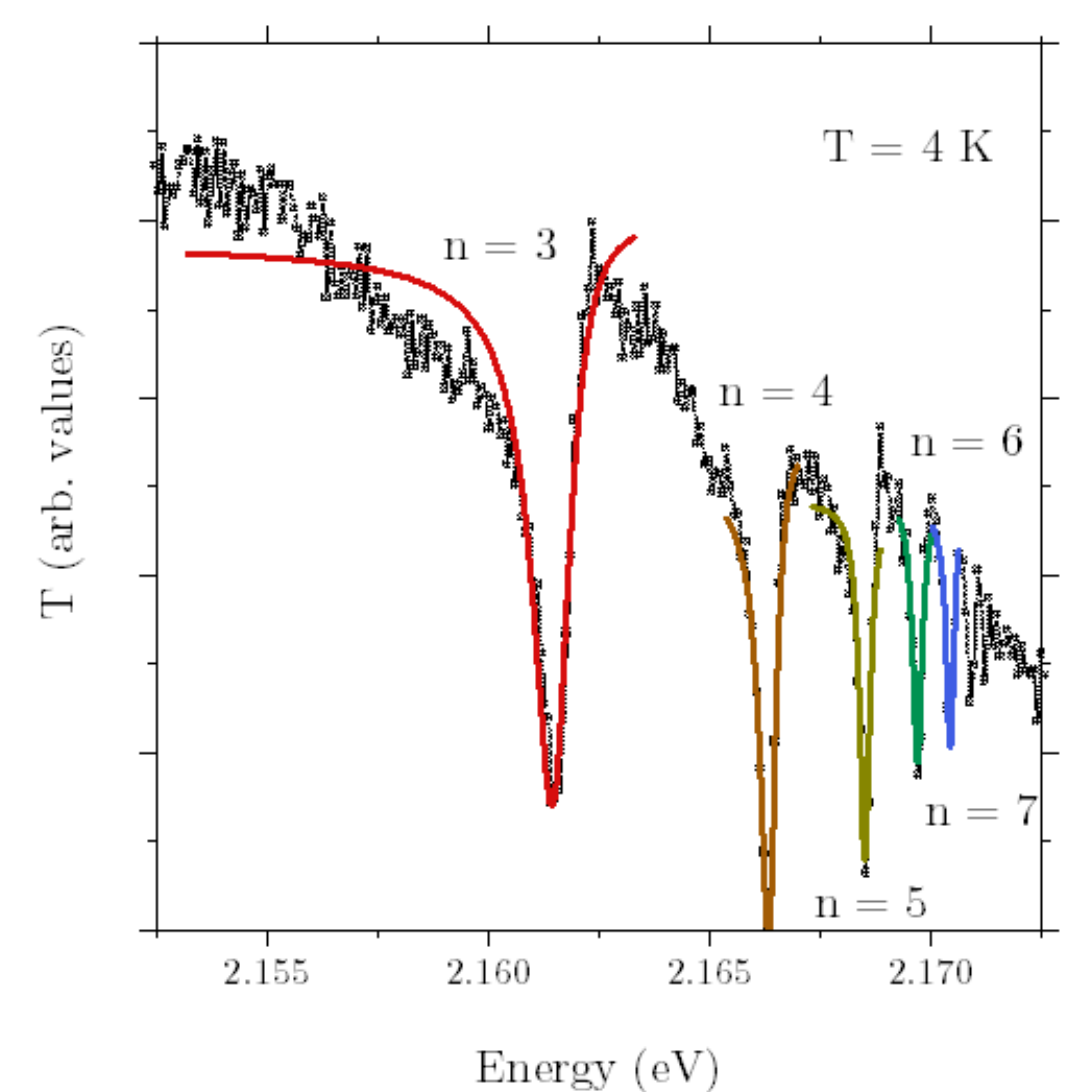}
    \caption{Transmission spectra of a bare {\cutwoo} flake. Points are experimental data. Solid curves are Fano lineshape fits for each resonance.}
    \label{fig:s2}
\end{figure}

From white light transmission spectra of a bare \cutwoo flake with excitonic resonances fitted as an asymmetrical Fano resonances we extract exciton linewidth $\gamma_n$ and Fano asymmetry parameter $Q_n$ (see Fig. \ref{fig:s2}).

\begin{figure}[H]
    \centering
    \includegraphics[width=0.55\columnwidth]{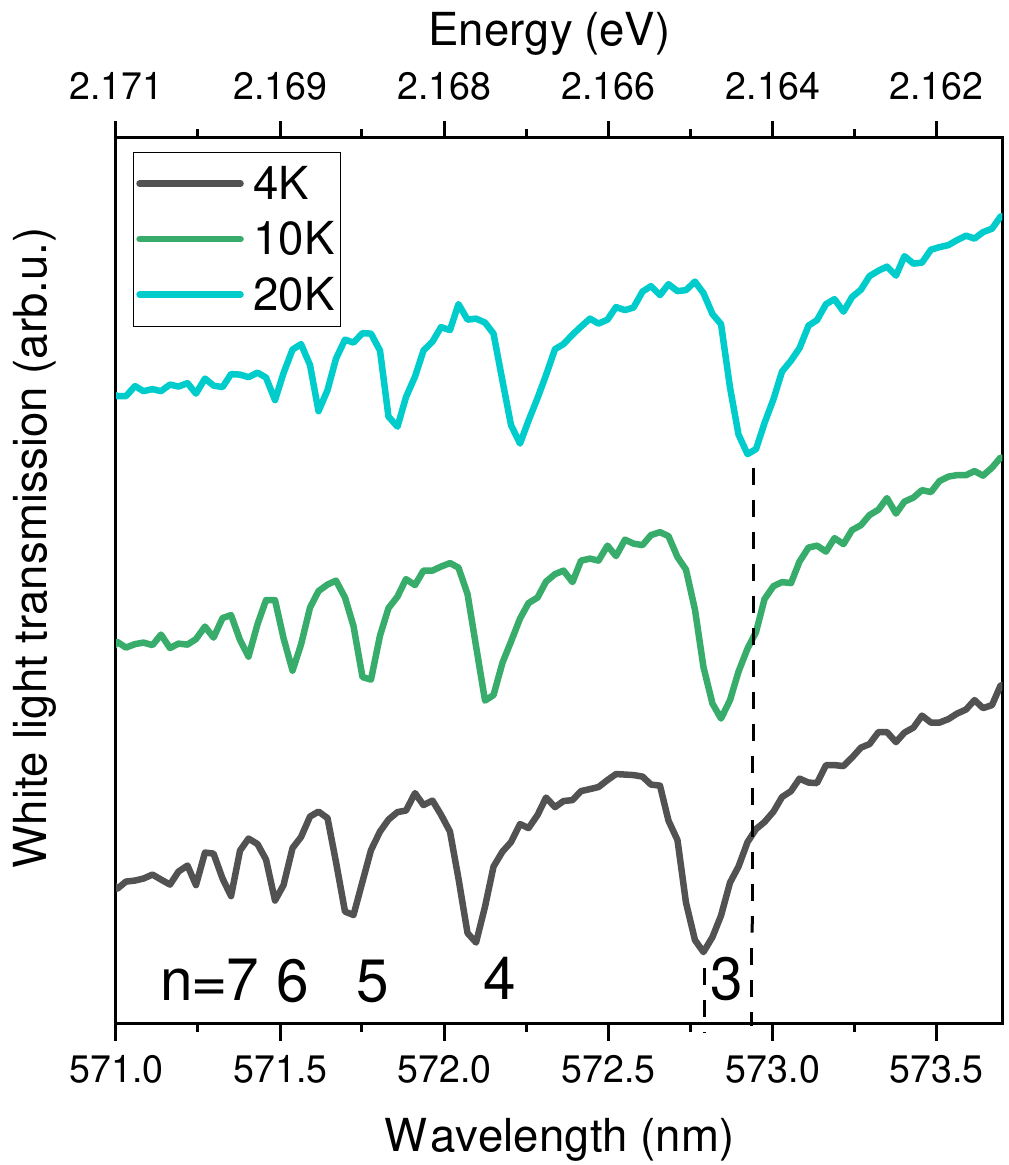}
    \caption{\textcolor{black}{White light transmission spectra of a bare {\cutwoo} flake at different temperatures. }}
    \label{fig:s4}
\end{figure}

\textcolor{black}{\section{The effect of temperature on the exciton resonances}}
\textcolor{black}{
We believe the laser induced heating effects are negligible in our experiment. To verify this we performed an additional measurement where the exciton resonances were monitored in transmission in a bare Cu$_{2}$O crystal using white light for different temperatures. With the increase of temperature of the sample from 4 to 20 K we observed a significant red shift of the exciton resonances by about 0.6 meV as shown in the Fig. \ref{fig:s4}, whereas the exciton linewidth and the dip of the exciton resonance stay almost the same, indicating no reduction of the exciton oscillator strength with temperature. By contrast, our measurements of  the polariton resonances for different powers in Fig. 2 of the main text do not show any red shift of the bare exciton resonances within the spectrometer resolution $\sim 0.1$ meV: the energy positions of the dip between the polariton resonances corresponding to the exciton levels do not change. This indicates that the sample temperature does not change with increase of pump power and the reduction of the exciton-photon coupling can not be explained by heating. 
}

\section{Strong coupling in broadband excitation regime up to n = 7}

In Fig. 1 of the main text we demonstrated strong coupling in the broadband excitation regime using angle-resolved transmission spectra, by showing the anti-crossing of the cavity modes around the excitonic lines for $n = 3$, $n = 4$, and $n = 5$. In this section, we show that such anti-crossing can be observed for up to $n = 7$ in the broadband excitation regime, similarly to the narrowband excitation scheme demonstrated in the main text of the article.

\begin{figure}
    \centering
    \includegraphics[width=\columnwidth]{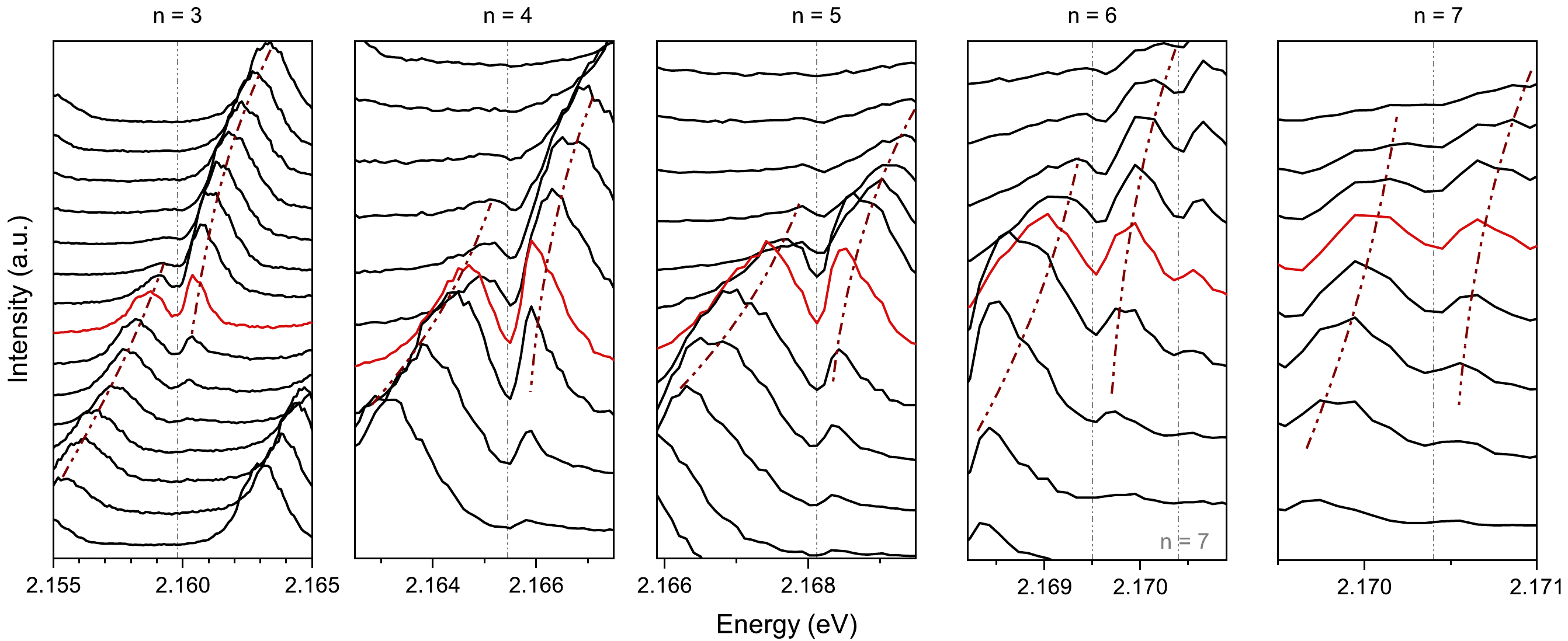}
    \caption{Broadband excitation and position scan across the sample showing anticrossing for excitonic resonances from $n = 3$ to $n = 7$. Brown dash-dot lines are a guide to the eye showing positions of upper and lower polariton modes. Grey lines are showing positions of exciton resonances.}
    \label{fig:waterfall_splittings}
\end{figure}

While Fig. 1 of the main text was obtained using $\bm{k}$-space imaging, it was not possible to use this technique for higher $n$. Indeed, the cavity modes become noisy and harder to resolve at higher energy, which makes the identification of such modes challenging. 
Instead, the anti-crossing is obtained by scanning the excitation position on the sample, which results in the cavity modes shifting due to the slight wedge in sample thickness and thus changing their detuning with respect to the excitons. This is similar to the technique used in Ref.~\cite{Orfanakis2022}. When the modes cross the exciton resonance, a doublet and splitting characteristic of strong coupling are observed. This scan across different positions on the sample is shown in Fig. \ref{fig:waterfall_splittings}. 

\section{Calculation of the density}\label{SM:density}

The density of photons inside a cavity can be calculated from the power emitted by the cavity together with the rate of escape through the mirrors, which have a finite transmission coefficient. A detailed explanation of the principle can be found in the supplementary information of Ref.~\cite{Kuriakose2022}. In the present work, the power emitted by the cavity was measured directly, while the transmission of the mirrors was calculated using a transfer matrix model of the sample with calibrated parameters. Transfer matrix modelling is a standard technique which exactly solves Maxwell's equations for planar layer structures of the type we use in this work.
In our work we deduce the size of the nonlinearity at the lowest powers, in the limit where the interaction energy compared to the the other energy scales in the system, such as losses, are tending towards zero. We therefore use the reasonable approximation that in this limit the density of particles in the cavity can be deduced using a linear model of the cavity electromagnetic response.

The first step in the model was to calibrate the reflection, transmission and absorption of the silver mirrors. In the same deposition runs in which the mirrors were deposited onto the \cutwoo to form the cavity the same silver films were also deposited onto a glass substrate. We then measured the transmission and reflection of these silver films using the same laser as in the main experiments, and a commercial laser power meter. We measured reflection of $R=92\pm 3$\% and transmission of $T=2.0\pm 0.3$\%. The silver thickness of 51.5 nm was known from the deposition rate and time. We then used the transfer matrix method to model the reflection and transmission of the structure and found the silver refractive index that gave the measured transmission and reflection. We find real and imaginary parts of the refractive index $\text{n}=0.24\pm0.11$ and $k=3.78\pm0.10$ respectively.
Here the errors were obtained by propagating the errors in $R$ and $T$ using the numerically calculated gradients $\partial \text{n}/\partial T$, $\partial \text{n}/\partial R$, $\partial k/\partial T$, and $\partial k/\partial R$. For the error $\Delta \text{n}$ in $\text{n}$ we used the formula $\Delta \text{n} = \lvert\partial \text{n}/\partial T\rvert \Delta T + \lvert\partial \text{n}/\partial R\rvert \Delta R$, and similar for the error $\Delta k$ in $k$.
We note that our refractive index is also consistent with values from the literature~\cite{Ferrera2019}. We checked that our deduced values of density do not vary by more than a few percent with the real part of the refractive index of the silver mirrors.

Next we calibrated the \cutwoo layer properties. The background refractive index 2.7386 was taken from the literature~\cite{Morin2022}. We then modelled transmission through the cavity using the transfer matrix method and varied the thickness of the \cutwoo until the free spectral range between adjacent photonic modes matched that in the experiment. In this manner we found a cuprous oxide thickness of 26.0 micrometers. The imaginary part of the cuprous oxide refractive index, 0.000483, was then found so that the quality factor of the calculated modes was 2000 away from the excitonic resonances, in agreement with the experimental results. We checked that our deduced values of density do not depend (by more than a few percent) on the refractive index of the cuprous oxide provided that the free spectral range matched the experimental value.

Having calibrated both mirror and cavity properties we then used the transfer matrix model to calculate both the energy density inside the cavity and the Poynting flux (power per unit area) outside the cavity for a monochromatic incident laser, as in Ref.~\cite{Kuriakose2022}. Integrating the energy density over the cavity length and dividing by the flux gives $\tau'$, the inverse of the rate of tunnelling of photons through the mirror. We then have $P_{\mathrm{out}}=E_{\mathrm{cav}}/\tau'$ where $P_{\mathrm{out}}$ is the measured power coming out of the cavity and $E_{\mathrm{cav}}$ is the energy stored as photons inside the cavity.

We calculated $\tau'$ as a function of photon wavelength and found variation of only 0.4\% over the 1.75 meV ($0.467$ nm) energy range corresponding to the bandwidth of the spectra shown in the experiments. Since in the linear regime the time-varying fields inside the cavity may be viewed as a superposition of different frequency waves (by the Fourier transform) we can therefore assume that the ratio of stored energy to output power is the same for the pulsed case as for the monochromatic case. We arrive at $\tau'=14\pm 2$ ps. Here the uncertainty $\Delta \tau'$ in $\tau'$ was obtained by propagating the errors $\Delta \text{n}$ and $\Delta k$ in the silver mirror real and imaginary refractive indices respectively using $\Delta \tau' = \lvert\partial \tau'/\partial \text{n}\rvert \Delta \text{n} + \lvert\partial \tau'/\partial k\rvert \Delta k$. The gradients $\partial \tau'/\partial \text{n} =3.27$ ps and $\partial \tau^{'}/\partial k = 19.48$ ps were obtained numerically by running the transfer matrix simulation over a range of different refractive indices.

The number of particles in the cavity is related to the energy by $N = E_{\mathrm{cav}}/\left(\hbar\omega\right)$ where $\omega$ is the central angular frequency of the pulses. The areal density of photons $\rho_{\mathrm{photons}}=N/A$ can then be obtained using the effective area $A$ for the nonlinear interaction. This is given by~\cite{Carusotto_2010} $1/A=\int\int_{-\infty}^{\infty} I^2(x,y) dxdy$ where $I(x,y)$ is the normalised spatial intensity distribution inside the cavity, which in our case has a Gaussian shape. This takes account of the fact that the density varies over the Gaussian spot and so we measure a weighted average of more and less strongly interacting regions.

In a similar way our time-averaged measurement of the $\sim 1$ ps pulses coming from the cavity is an average over the time-varying interaction energy in the cavity weighted by the occupancy of the cavity. Both the interaction energy and occupancy are proportional to the temporal shape of the cavity occupancy $I(t)$.
Taking this weighted average gives the effective pulse length $\tau_{\mathrm{p}}$ using $1/\tau_{p} = \int_{-\infty}^{\infty} I^2(t) dt$.
Transform limited Gaussian pulses with 1.75 meV spectral width have a temporal width of 1.0 ps. This is then a lower bound for the possible temporal width of the density in the cavity, which provides an upper bound for the density and hence a lower bound for the nonlinearity. However, the real temporal width is likely to be longer due to partial cavity filtering of the incident spectrum and/or chirp in the incident pulses. Using the lower bound of a 1 ps Gaussian we obtain $\tau_{\mathrm{p}}=1.57$ ps.

We then use $E_{\mathrm{cav}}=P_{\mathrm{out}}\tau'$ from above and insert the effective output pulse power $P_{\mathrm{out}}=P_{\mathrm{avg,T}}/\left(f\tau_{\mathrm{p}}\right)$. Here $P_{\mathrm{avg,T}}$ is the average power of the \textit{transmitted} beam, $f=1$ kHz is the laser repetition rate and $\tau_{\mathrm{p}}$ is the effective pulse length defined above. We finally make the substitution $P_{\mathrm{avg,T}}=T P_{\mathrm{avg}}$ where $P_{\mathrm{avg}}$ is the average \textit{incident} beam power and $T=1/180$ is the measured transmission through the cavity. This accounts for the reflection of spectral components of the incident pulse which are not resonant with the cavity modes. Recalling that $\rho_{\mathrm{photons}}=N/A=E_{\mathrm{cav}}/\left(A\hbar\omega\right)$ finally leads to an expression for the photon density in cavity:
\begin{equation}\label{eq:photon_density}
    \rho_{\mathrm{photons}} =\frac{T P_{\mathrm{avg}}\tau'}{f \tau_\mathrm{p} A \hbar \omega} .
\end{equation}

So far our discussion has concerned a purely photonic cavity with no strong exciton-photon coupling. When strong coupling is included the fundamental eigenstates of the system, the polaritons, are part photon and part exciton. The fractions of photon and exciton content are $\lvert C\rvert^{2}$ and $\lvert X\rvert^{2}$ respectively with $\lvert C\rvert^{2}+\lvert X\rvert^{2}=1$. Since our measurements of nonlinearity are made at zero exciton-photon detuning we have $\lvert C\rvert^{2}=\lvert X\rvert^{2}=0.5$. Only the photonic component of the polaritons leads to tunnelling through the mirror into free space modes outside the cavity. Thus Eqn.~\ref{eq:photon_density} still holds and the output power from the cavity gives the density of photons inside the cavity, where we are careful to remember that this is really the density of the photonic component of the polaritons. The density of the photonic component is related to the density of polaritons by $\rho_{\mathrm{photons}} = \rho_{\mathrm{polaritons}}\lvert C\rvert^{2}$. We can then write the density of polaritons as
\begin{equation}
    \rho_{\mathrm{polariton}} =\frac{T P_{\mathrm{avg}}\tau'}{f \tau_\mathrm{p} A \hbar \omega \lvert C\rvert^{2}} .
\end{equation}

Finally, to obtain the density of excitons (the excitonic component of the polaritons) we multiply the polariton density by the excitonic fraction,
\begin{equation}
    \rho =\frac{T P_{\mathrm{avg}}\tau'\lvert X\rvert^{2}}{f \tau_\mathrm{p} A \hbar \omega \lvert C\rvert^{2}}.
\end{equation}

\section{Comparison of nonlinearities}

We find that in our system the nonlinearity coefficient $\beta$, relevant for applications, ranges between 0.01~$\mu$eV $\mu$m$^3$ for $n=3$, to 0.4~$\mu$eV $\mu$m$^3$ for $n=7$ Rydberg exciton-polaritons (see Fig.~3 in the main text). It is important to note that in a Cu$_2$O cavity the excitons are delocalised within the cavity thickness of $26~\mu$m, and so the exciton density is expressed per unit volume of the cavity region and the appropriate units for the nonlinear parameter are energy shift divided by number of particles per unit volume, that is $\mu$eV $\mu$m$^3$.

In other highly nonlinear polariton systems, such as for example microcavities with embedded (In)GaAs quantum wells, the excitons are confined within the thickness of the quantum wells (typically $\sim$ 10 nm per quantum well in the device). A single quantum well thickness is comparable to the exciton Bohr radius and hence the density is usually expressed per unit area of a single quantum well. For GaAs polaritonic systems the reported strengths of exciton-polariton nonlinearity (either $\beta$-values or $g$-values characterising the collapse of strong exciton-photon coupling or the exciton energy shifts, respectively) are in the range from 2 to 10~$\mu$eV $\mu$m$^2$ \cite{Walker2015, Kuriakose2022, Ferr2011,EstrPRB2019,Brichkin2011}. In order to compare these to $\beta$-values of the bulk excitons we study, one has to transform the 2D density to effective 3D density by dividing it by the thickness of the quantum wells.
This enables a unified characterisation of the strength of interactions between two excitons separated by a certain distance irrespective of how they are positioned within the cavity region, whether they are bulk or confined to a single or multiple 2D layers.

Following this procedure the 2D values of 2--10~$\mu$eV $\mu$m$^2$ in GaAs-based systems are equivalent to 0.02--0.1~$\mu$eV $\mu$m$^3$. These values are exceeded by the $\beta$ values in \cutwoo microcavity already for $n = 5$ exciton-polaritons. Qualitatively, this is expected since the exciton Bohr radius for $n=5$ is already of the order of 30 nm, being three times larger than that in GaAs, leading to stronger dipole-dipole interactions or Pauli blockade mechanism. 
Similarly, nonlinearities in hybrid perovskites containing order 3000 layers have been studied~\cite{Fieramosca_2019}. Each layer is of order 1.7 nm thick. The per-layer nonlinearity of 3~$\mu$eV $\mu$m$^2$ is equivalent to a bulk-like nonlinearity of 0.005~$\mu$eV $\mu$m$^3$. This 3D value is then convenient to deduce the effective 2D nonlinearity of perovskite structures with different numbers of layers.

\section{Theoretical analysis}

In this section we present the theoretical analysis of Rydberg excitons coupled to photons in a microcavity. In the first part, we focus on the relation of the Rabi splitting ($\Omega_n$) and the light-matter coupling constant ($G_n$). In particular, we derive $\Omega_n \approx \Omega_n^{(0)} - \beta_{n} \rho$, where $\rho$ is the exciton density, and $\Omega_n^{(0)}$ is the Rabi splitting at vanishing density, $\rho=0$. This gives the theoretical beta factor $\beta_n$ used to characterise the strength of nonlinearity. In general, $\beta_{n} \sim V_{\mathrm{B}}/V$ is mostly determined by the ratio between blockade ($V_\mathrm{B}$) and the total volume ($V$). In the next part, we discuss the blockade $V_\mathrm{B}$ due to Rydberg and Pauli blockade. These two different mechanisms lead to distinct scaling behavior. Also, we discuss the estimates for $\beta_n$ in Cu$_2$O (Fig.~3\textbf{b} in the main text). In the final part, we deduce the nonlinear refractive index n$_2$ for Cu$_2$O using the nonlinear polaritonic response.

\subsection{Rabi splitting and light-matter coupling}\label{SM:Rabi}

Transmission of the Cu$_2$O microcavity system around $n$-th excitonic state can be modeled as \cite{Orfanakis2022}
\begin{equation}\label{eqn:Tn}
    T_n\approx\frac{1}{(\frac{\kappa}{2}+G_n^2\frac{\gamma_n/2+2Q_n\Delta_n}{\gamma_n^2/4+\Delta_n^2})^2+(\Delta_n-G_n^2\frac{\Delta_n-Q_n\gamma_n}{\gamma_n^2/4+\Delta_n^2})^2},
\end{equation}
where $\kappa$ is the cavity line-width, $G_n$ is the exciton-photon coupling constant, $\Delta_n$ is the detuning, $\gamma_n$ is the excitonic linewidth, and $Q_n$ is the Fano asymmetry parameter. In the weak light-matter coupling regime ($G_n \ll\gamma_n$), the system only responds to light with frequency near the exciton resonance ($\Delta_n=0$). One can see from Eq. \eqref{eqn:Tn} that in the strong coupling regime ($G_n \gg\gamma_n$) the optical response changes qualitatively \cite{Khitrova:RMP71(1999)}. Namely, in this regime the resonance changes from $\Delta_n=0$ to two resonances at $\Delta_n=\pm\frac{1}{2}\Omega_n$. This comes from the hybridisation of photonic and excitonic modes. The resulting states -- polaritons -- are quasiparticles which energies are characterised by the Rabi splitting $\Omega_n$.

The Rabi splitting can be analytically calculated from Eq.~\eqref{eqn:Tn} by identifying the separation between points of maximum response in the transmission spectrum. For instance, in the absence of asymmetry ($Q_n=0$), Rabi splitting can be obtained from Eq. \eqref{eqn:Tn} as \cite{Savona:SSComm93(1995)}  
\begin{equation}\label{eqn:Omegan}
    \Omega_n=2\sqrt{G_n\sqrt{\tfrac{1}{2}\gamma_n(\kappa+\gamma_n)+G_n^2}-\tfrac{1}{4}\gamma_n^2} .
\end{equation}
We can see that the Rabi splitting explicitly depends on the coupling constants $G_n$ and the linewidth $\gamma_n$. These quantities can be exciton density-dependent. For instance, the exciton blockade can lead to the reduction of $G_n$, and the scattering between excitons broadens the linewidth $\gamma_n$. These effects will eventually renormalise the Rabi splitting or the shift of polariton energy which directly translates into optical nonlinearity. To see these effects, we expand Eq.~\eqref{eqn:Omegan} at low exciton density $\rho$ as 
\begin{equation}\label{eqn:Rabi}
    \Omega_n=\Omega_n^{(0)}-\beta_n \rho +\mathcal{O}(\rho^2),
\end{equation}
where the Rabi splitting in low density is 
\begin{equation}\label{eqn:Omega0}
    \Omega_n^{(0)}=2\sqrt{G_n^{(0)}\Lambda_n-\tfrac{1}{4}(\gamma_n^{(0)})^2} ,
\end{equation}
with $G^{(0)}_n :=G_n\vert_{\rho=0}$, $\gamma^{(0)}_n :=\gamma_n\vert_{\rho=0}$, and $\Lambda_n :=\sqrt{\frac{1}{2}\gamma_n^{(0)}(\kappa+\gamma_n^{(0)})+(G^{(0)}_n)^2}$. Here, the $\beta$-factor then reads
\begin{align}\label{eqn:betan}
    \beta_n=-\frac{2}{\Omega_n^{(0)}\Lambda_n}\Big[&(\Lambda_n^2+(G_n^{(0)})^2)\frac{d G_n}{d \rho}+G_n^{(0)}(\tfrac{1}{2}\kappa+\gamma^{(0)}_n)\frac{d\gamma_n}{d \rho}\Big].
\end{align}
This factor quantifies the rate of the Rabi splitting reduction. In the above, the exciton blockade and the linewidth broadening effects are present in the first and second terms. However, no strong inhomogeneous broadening has been resolved in the measurement within the low-density regime. Therefore, in our analysis, we focus on the blockade effect in the first term. 

%-- Sec. 6.2

\subsection{Rydberg and Pauli blockade}\label{SM:blockade}
\begin{figure}
    \centering
    \includegraphics[width=\columnwidth]{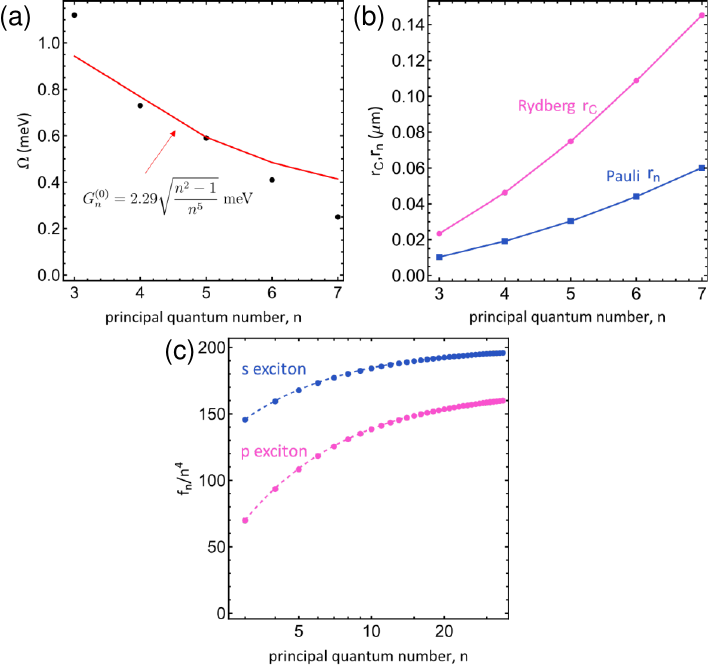}
    % \caption{Scaling properties of Rydberg and Pauli exciton. (a)Coupling strength in low density. Closed blue circles are the point taking from Fig. 3.a of the main text at $\rho=0$. Dashed curved is the fitted result by using $G_n^{(0)}\propto\sqrt{(n^2-1)/n^5}$ and Eq. \eqref{eqn:Omegan}.}
    % \label{fig:Omega_fit}
     \caption{Scaling of Rydberg exciton. (a) Rabi splitting in low density. Closed black circles are the point taking from Fig.~3\textbf{a} of the main text at $\rho=0$. Dashed curved is the fitted result by using $G_n^{(0)}\propto\sqrt{(n^2-1)/n^5}$ and Eq.~\eqref{eqn:Omegan}. (b) Comparison between Rydberg ($r_C$ magenta) and Pauli ($r_n$, light blue) blockade radius. (c) Scaling of Pauli blockade coefficient. The factor $f_n$ in Eq.~\eqref{eqn:f2} gives the asymptotic scaling behavior for the blockade coefficient $B_n\sim n^4$  [Eq.~\eqref{eqn:sigma2}]. Dashed curves are the fits by using $f_n=n^4(b_0+b_1n^{-1}+b_2n^{-2})$.} %$f_n=n^4(b_0+b_1n^{-1}+b_2n^{-2})$ with $b_0=201$, $b_1=165$, $b_2=0$ for $s$ exciton and $b_0=169$, $b_1=32$, $b_2=308$ for $p$ exciton
    \label{fig:Scaling}
\end{figure}

In this subsection, we discuss the possible blockade mechanism that leads to reduction of $G_n$, and present the details for derivations. First, let us comment on the case of Rydberg excitons outside of optical cavities. In the presence of $N$ Rydberg excitons, the absorption ($\alpha$) of Cu$_2$O follows the scaling relation $\alpha\propto V/V_\mathrm{B}-N$, where $V$ is the total volume of the system and $V_\mathrm{B}$ is the Rydberg blockade volume~\cite{Kazimierczuk2014}. The $N$-dependent behavior in $\alpha$ can be well explained by Rydberg blockade physics~\cite{Amann2020}. When placed inside an optical cavity, the absorption is related to the light-matter coupling constant as $\alpha\propto G_n^2$ \cite{Orfanakis2022}. In the low-density limit, the coupling constant may be written as 
\begin{equation}\label{eqn:Gn}
    G_n\approx G_n^{(0)}(1-\tfrac{1}{2}B_n N),
\end{equation}
where $B_n=V_\mathrm{B}/V$ is the blockade coefficient for a single Rydberg exciton. In the case of Rydberg blockade it is given by
\begin{equation}\label{eqn:Bn-Rydberg}
    B_n=\frac{4\pi}{3} \frac{r_C^3}{V}. \quad(\text{Rydberg})
\end{equation}
The Rydberg exciton blockade radius is modeled by $r_C=(C_k/\gamma_n)^{1/k}$ with $C_k$ being the dipole-dipole interacting constant~\cite{Amann2020,Walther:PRB98(2018)}. Here, $k=3$ is the F\"{o}rster-type interaction and $k=6$ is the van der Waals interaction. This parameter plays a crucial role in determining blockade physics which has been estimated theoretically in Ref. \cite{Walther:PRB98(2018)}. To calculate the $\beta$-factor, we substitute Eq. \eqref{eqn:Gn} into Eq. \eqref{eqn:betan}, and get
\begin{equation}\label{eqn:betan-blockade}
    \beta_n=\frac{\Lambda_n^2+(G^{(0)}_n)^2}{\Omega_n^{(0)}\Lambda_n}G^{(0)}_nB_nV. %A
\end{equation}
%
%where we have let $\rho=N/A$ with $A$ being the area of the sample. 
We then extract the light-matter coupling constant $G_n^{(0)}$ from the measurement in Fig.~3\textbf{a} of the main text by using Eq.~\eqref{eqn:Omega0}. This gives the best fit $G_n^{(0)}=2.29(\frac{n^2-1}{n^5})^{1/2}$meV in the main text, see Fig.~\ref{fig:Scaling}\textbf{a}. %Assuming $V=L A$ with $L=30\mu$m being the thickness of the sample (see Fig. 1a), 
We obtain the $\beta$-factor plotted as the purple solid curve in Fig.~3\textbf{b}.
%%%
%%%
In the large $n$-limit, the blockade coefficient shows a power-law scaling with $B_n\sim n^{7}$\cite{Amann2020}. Hence, Eq. \eqref{eqn:betan-blockade} gives the asymptotic scaling for $\beta_n\sim n^{5.5}$.

However, this scaling property which is very often used for identifying the Rydberg blockade has only been established in the large quantum number regimes ($n\geq12$)~\cite{Amann2020}. In our case, the exciton quantum number are in the range from $n=3$ to $n=7$, where the scaling behavior may not be evident. Therefore, the power-law scaling may not be a single argument supporting the observation of the Rydberg blockade in our low-$n$ measurement. Next, we consider another potential contribution to the reduction of Rabi frequency.

As the exciton radius $r_n=\tfrac{1}{2}a_0(3n^2-2)$ \cite{Amann2020} is comparable to the Rydberg blockade radius $r_C$  [see Fig.~\ref{fig:Scaling}(b)], we consider the effects of Pauli blockade. This comes from the composite nature of excitons and fermionic statistics of the electrons and holes. In order to identify the Rydberg physics in this low-$n$ regime, we analyse the Pauli blockade and investigate its contribution. 

For the Pauli blockade, the coefficient in Eq.~\eqref{eqn:Gn} can be evaluated exactly as 
\begin{equation}
 B_n=\sum_{\bm{k}}\vert\psi_n(\bm{k})\vert^4, \quad(\text{Pauli}),
\end{equation}
where $\psi_n(\bm{k})$ is the exciton wavefunction with quantum number $n$ and wavevector $\bm{k}$ \cite{Combescot2008}.
The $p$-wave hydrogen-like wavefunction \cite{Podolsky:PR34(1929)} ($l=1, m=0$) is
\begin{align}
    \psi_n(\bm{k})=&\sqrt{\frac{(n a_0)^3}{V}\frac{n(n-l-1)!}{(n+l)!}}\Big[
    C^{(l+1)}_{n-l-1}(\tfrac{\xi^2-1}{\xi^2+1})\frac{2^{2l+3}2\pi l!\xi^l}{(\xi^2+1)^{l+2}}\Big]\Upsilon_{lm}(\theta,\phi) ,
\end{align}
where $\xi= n k a_0$. Here, $C^{(\alpha)}_n(x)$ is the Gegenbauer functions and $\Upsilon_{lm}(\theta,\phi)$ is the spherical harmonic function. We note that the momentum $\bm{k}$ is discrete, defined by a finite sample size with volume $V$. The normalisation condition is $\sum_{\bm{k}}\vert\psi_n(\bm{k})\vert^2=1$.
In contrast to the Rydberg blockade, the shape of the exciton wavefunction $\psi_n(\bm{k})$ completely determines $B_n$ or the Bohr radius $a_0$. The Bohr radius can be determined by the experiment's Rydberg exciton energies, $\omega_n=E_g+E_n^b$ in Fig.~\ref{fig:s2} with bandgap energy $E_g$ and the exciton binding energy 
\begin{equation}
    E_n^{b}=-\frac{\mathrm{Ry}^\ast}{n^2}.
\end{equation}
The Rydberg constant $\mathrm{Ry}^\ast=\frac{e^2}{ 4\pi \epsilon_0\epsilon_r}\frac{1}{2a_0}$, which allows us to estimate $a_0$.
Using the Cu$_2$O dielectric constant $\epsilon_r=7.5$ \cite{Kavoulakis:PRB55(1997)}, and  
\begin{equation}\label{eqn:Ry-n}
    \mathrm{Ry}^\ast=-\frac{\omega_{n-1}-\omega_{n}}{(n-1)^{-2}-n^{-2}},
\end{equation}
we can deduce the Bohr radius $a_0\approx0.83$~nm. Therefore, in the Pauli blockade, the experiment leaves no free adjustable parameter for the $\beta$-factor. We plot $\beta_n$ from the contribution due to the Pauli blockade in Fig. 3 of the main text (blue dashed curve). It is an order of magnitude smaller than the measured values. 

In terms of power-law scaling, we let
\begin{equation}\label{eqn:sigma2}
    B_n=f_n\frac{a_0^3}{V}.\quad \text{(Pauli)}
\end{equation}
The prefactor is
\begin{equation}\label{eqn:f2}
    f_n=\frac{9\times2^{20}n^3}{10(n^2-1)^2}\int_0^{\infty}d\xi \frac{\xi^6[C^{(2)}_{n-2}(\tfrac{\xi^2-1}{\xi^2+1})]^4}{(1+\xi^2)^{12}},
\end{equation}
where we used $\sum_{\bm{k}}\frac{(2\pi)^3}{V}\to\int d^3k$ for wavevector $\bm{k}$ in large $V$. This $f_n$ prefactor determines the scaling behavior of $B_n$ and we plot in Fig. \ref{fig:Scaling}(c). As we can see, the Pauli blockade coefficient ($B_n$) scales with a power law weaker than $n^4$ for low $n$, and approaches an $\sim n^{3.5}$ scaling in the high-$n$ range.
%with lower power as $n^5$ in low-$n$ with asymptotic scaling as $n^4$.  
Overall, it is lower than the Rydberg blockade scaling with $n^7$. Using Eq. \eqref{eqn:betan-blockade}, this yields a a scaling trend of $\beta_n\sim n^{2.5}$ for small $n$, significantly different as compared to experiment.

\subsection{n$_2$ parameter from Rabi frequency measurement}\label{SM:n2-theory}
To estimate the n$_2$ nonlinear parameter, we begin with the definition of the total refractive index and the optical susceptibility as follows \cite{Morin2022}
\begin{equation}
    \mathrm{n_T}^2=\epsilon_b+\chi(\omega).
\end{equation}
The optical susceptibility in a cavity for each excitonic mode near the resonance can be modeled by \cite{Orfanakis2022,Khitrova:RMP71(1999)} 
\begin{equation}\label{eqn:chi}
    \chi(\omega)\approx\frac{h_n G_n^2}{\omega-\omega_n+\frac{1}{2}i\gamma_n},\quad(\omega\simeq\omega_n) ,
\end{equation}
where $\omega_n$ is the Rydberg exciton energy. We note that the constant of proportionality $h_n$ can be determined from the Rabi splitting measurement.

The coupling constant $G_n$ changes due to the Rydberg blockade as the laser power increases [Eq. \eqref{eqn:Gn}]. This leads to the nonlinear response in the susceptibility 
\begin{equation}
    \chi(\omega)=\chi^{(1)}(\omega)+\chi^{(3)}(\omega)E^2.
\end{equation}
In the vicinity of $\omega\simeq\omega_n$, the linear response of the above is 
$\chi^{(1)}(\omega)\approx h_n(G_n^{(0)})^2(\omega-\omega_n+\frac{1}{2}i\gamma_n)^{-1}$, and the Kerr nonlinear response is
\begin{equation}\label{eqn:chi3_E}
    \chi^{(3)}(\omega)\approx\frac{h_n( G_n^{(0)})^2(-B_n)(\frac{1}{2}\epsilon_0V/\omega)}{\omega-\omega_n+\frac{1}{2}i\gamma_n},
\end{equation}
where we have converted the exciton number $N$ into the electric field $E$ by using $\omega N/V\approx\frac{1}{2}\epsilon_0 E^2$ with $V$ being the volume of the nonlinear medium. Also, the blockade coefficient is given by Eq. \eqref{eqn:Bn-Rydberg}. The nonlinear refractive index n$_2$ is defined as
\begin{equation}
    \mathrm{n}_2(\omega)=\frac{\mathrm{Re}[\chi^{(3)}(\omega)]}{\epsilon_0 c \mathrm{n}_0^2},
\end{equation}
where $\mathrm{n}_0^2=\epsilon_b+\mathrm{Re}[\chi^{(1)}(\omega)]$. Therefore, the n$_2$-parameter (near $\omega\simeq\omega_n$) from the quench of Rabi frequency is 
\begin{equation}\label{eqn:n22}
    \mathrm{n}_2(\omega)\approx-\frac{ h_n  \beta_n }{2 c \mathrm{n}_0^2 \omega }\frac{G_n^{(0)}(\omega-\omega_n)}{(\omega-\omega_n)^2+\frac{1}{4}\gamma_n^2} ,
\end{equation}
where we have used Eq.~\eqref{eqn:betan-blockade} by taking the strong-coupling limit ($G_n^{(0)} \gg \gamma_n^{(0)}$). 

At the polaritonic peaks ($\omega=\omega_n\pm\tfrac{1}{2}\Omega_n^{(0)}$), the total refractive index satisfies~\cite{Khitrova:RMP71(1999)}
\begin{equation}\label{eqn:nT-Rabi}
    \mathrm{n_T}(\omega_n\pm\tfrac{1}{2}\Omega_n^{(0)})= \mathrm{n_b}\frac{\omega_n}{\omega_n\pm\tfrac{1}{2}\Omega_n^{(0)}}
\end{equation}
where $\mathrm{n_b}=\mathrm{n_T}(\omega_n)=\sqrt{\epsilon_b}$ is the background refractive index ($\mathrm{n_b}=1$ for the device is vacuum). The condition in Eq. \eqref{eqn:nT-Rabi} determines the constant of proportionality $h_n$. Therefore, we can deduce the nonlinear refractive index n$_2$ from our Rabi splitting measurements.

{\color{black} Alternatively, we can also deduce the constant of proportionality $h_n$ from the absorption data by following the method in Ref. \cite{Khitrova:RMP71(1999)}. First, the Rabi splitting of a nonlinear medium with susceptibility in Eq. \eqref{eqn:chi} can also be estimated as
\begin{equation}
    \Omega_n^{(0)}=\sqrt{2\omega_n h_n(G^{(0)}_n)^2/n_b^2-(\gamma_n^{(0)})^2}
\end{equation}
Substituting the experimental measurements into the above, we get $h_n$ which has the same orders of magnitude as the $h_n$ obtained from Eq. \eqref{eqn:nT-Rabi}. Furthermore, in Ref.\cite{Khitrova:RMP71(1999)}, we have $h_n(G^{(0)}_n)^2\approx(n_bc\alpha_0\gamma_n)/(\omega_n/\hbar)$ where $\alpha_0$ is the absorption. Using the absorption data in Ref. \cite{Kazimierczuk2014}, we again obtain the $h_n$ with the same order of magnitude.}

\textcolor{black}{\section{Non-resonant pumping and quenching of Rabi Splitting}}

\begin{figure}[!htb]
    \centering
    \includegraphics[width=0.55\columnwidth]{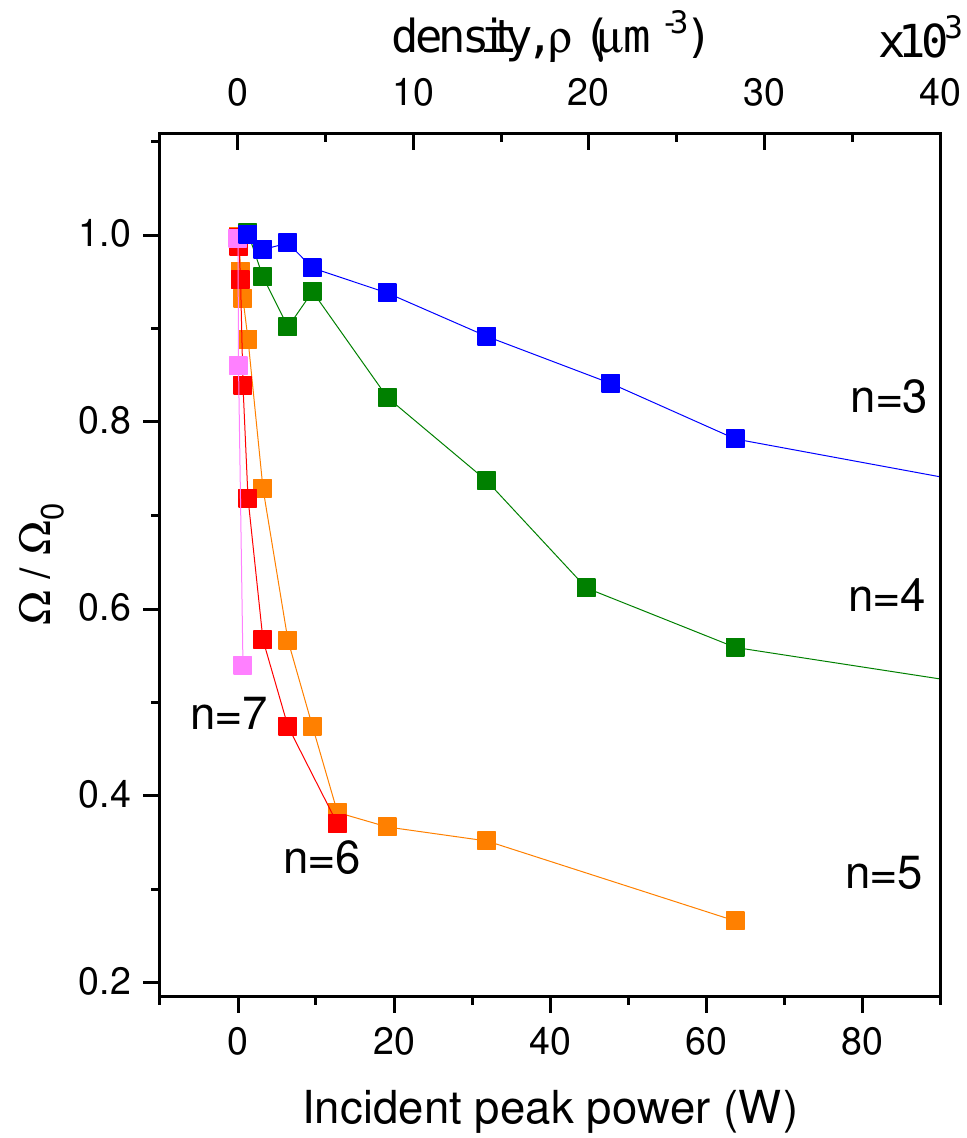}
\caption{\textcolor{black}{Normalized Rabi splitting as a function of resonant peak laser power (bottom axis) and/or photon density in the cavity (top axis). Normalization is based on the case where smallest value of laser power is used. }}
    \label{Fig:OmegaPeak}
\end{figure} 

\begin{figure}[!htb]
    \centering
    \includegraphics[width=0.58\columnwidth]{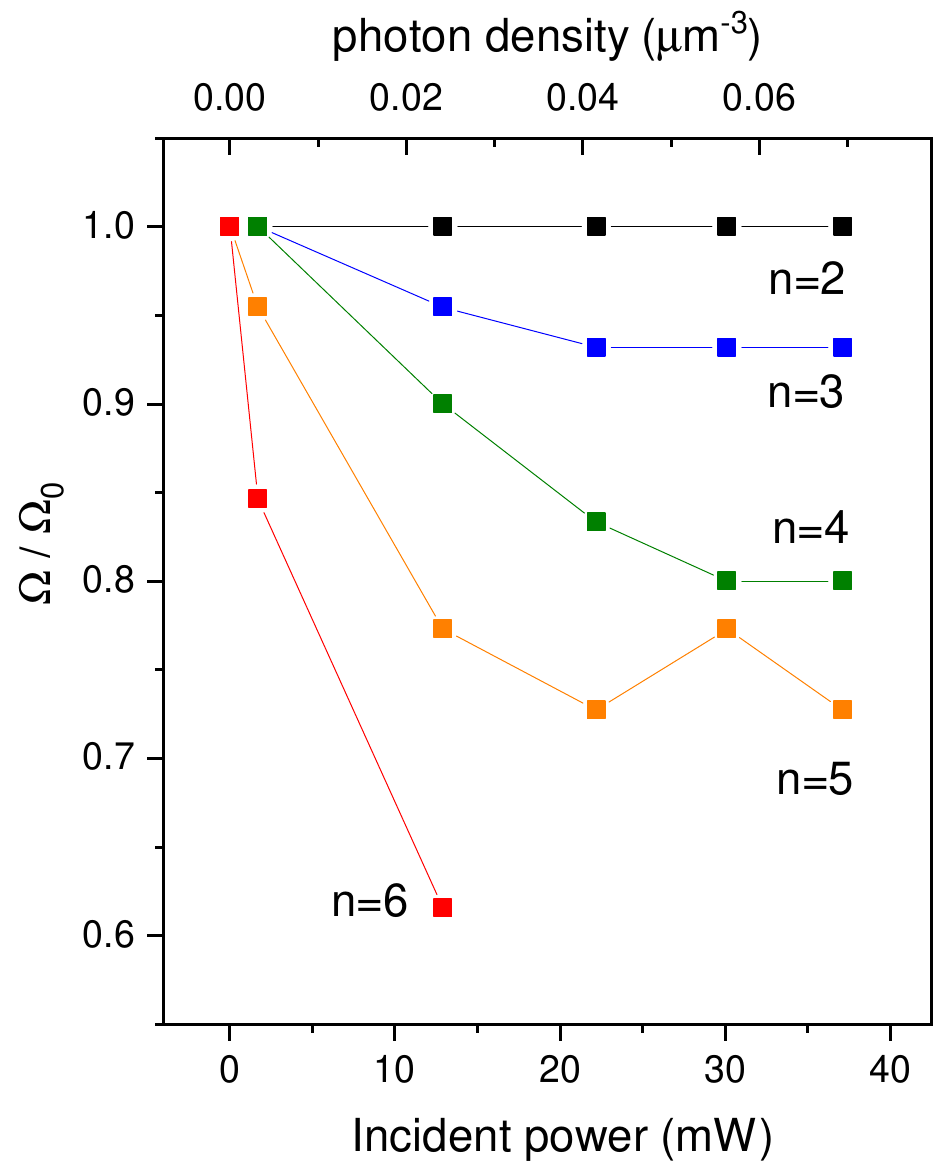}
    \caption{\textcolor{black}{Normalized Rabi splitting as a function of non-resonant CW laser power (bottom axis) and/or photon density inside the cavity (top axis). Rabi splitting values are normalized based on the case where no non-resonant laser is used. }}
    \label{fig:Fig5SH}
\end{figure}

\textcolor{black}{In this section we compare two regimes of quenching of Rabi splitting. One case is the resonant pulsed excitation presented in the main paper in Fig. 3. Here we show these data as normalised Rabi splitting plotted as a function of incident peak power and/or photon density created by the pulse in the active cavity region (Fig.~\ref{Fig:OmegaPeak}). The second case is non-resonant continuous wave (CW) excitation with above band gap green laser at wavelength 520 nm (Fig.~\ref{fig:Fig5SH}). Varying the power of non-resonant laser and probing with weak broadband super-continuum laser filtered from 568 to 582 nm in transmission geometry the transmission spectra are recorded and then fitted to extract Rabi splitting. Normalised Rabi splitting for this experiment is plotted as a function of CW power of green laser or photon density in the cavity for excitons from $n=2$ to $6$ (see Fig.~\ref{fig:Fig5SH}).The photon density in pulsed case is calculated using Eq.~\ref{eq:photon_density} whereas photon density in CW case is calculated using the following equation: 
\begin{equation}\label{eqn:PhotonDensityCW}
    \rho^{CW} = \tfrac{TP_{inc}}{3A\hbar\omega(c_0/n_0) },
\end{equation}
%$\upmu$J cm$^{-2}$
where T is the transmission through the mirror(2$\%$), $P_{inc}$ is the incident power, A is the illuminated area (95 $\upmu$$m^2$), $\hbar\omega$ is the photon energy, $c_0$ is the speed of light in vacuum, $n_0$ is Cu$_2$O background refractive index, 1/3 is from the fact that propagation length of the excitation laser $\approx$ 10 $\upmu$m in Cu$_2$O with thickness of $\approx$ 30 $\upmu$m. The data for CW case are obtained on microcavity \cutwoo sample with distributed Bragg reflector mirrors reported in Ref.~\cite{Orfanakis2022}, which is very similar in optical characteristics to the sample presented in the main paper --- cavity with silver mirrors. Note that the incident powers used in CW excitation case are in tens of mW range for quenching of Rabi splitting whereas in resonant experiment with pulsed excitation quenching of Rabi splitting is achieved at peak powers of tens of Watts. Even higher difference can be noted in the photon density required for quenching the Rabi between CW and pulsed regime (6 orders of magnitude). Such big difference of 6 orders of magnitude in photon densities can be explained by population of free electrons and holes and long lived states in the case of CW excitation contributing to nonlinear behaviour and quenching of Rabi splitting. We also note that in our experiments the photon densities in case of non-resonant CW pumping are only 6 orders of magnitude less than in the case of pulsed excitation. So our experiment with CW nonresonant pumping alone cannot explain why there is 8 orders of magnitude difference in the n$_2$ parameters measured in the case of resonant pulsed and CW pumping in Ref. \cite{Morin2022}. It is possible that such a difference is sample dependent (for example, the density of long-lived localised states, which could be associated with metallic impurities may vary from sample to sample).}

%\newpage

\begin{figure}[!htb]
    \centering
    \includegraphics[width=0.95\columnwidth]{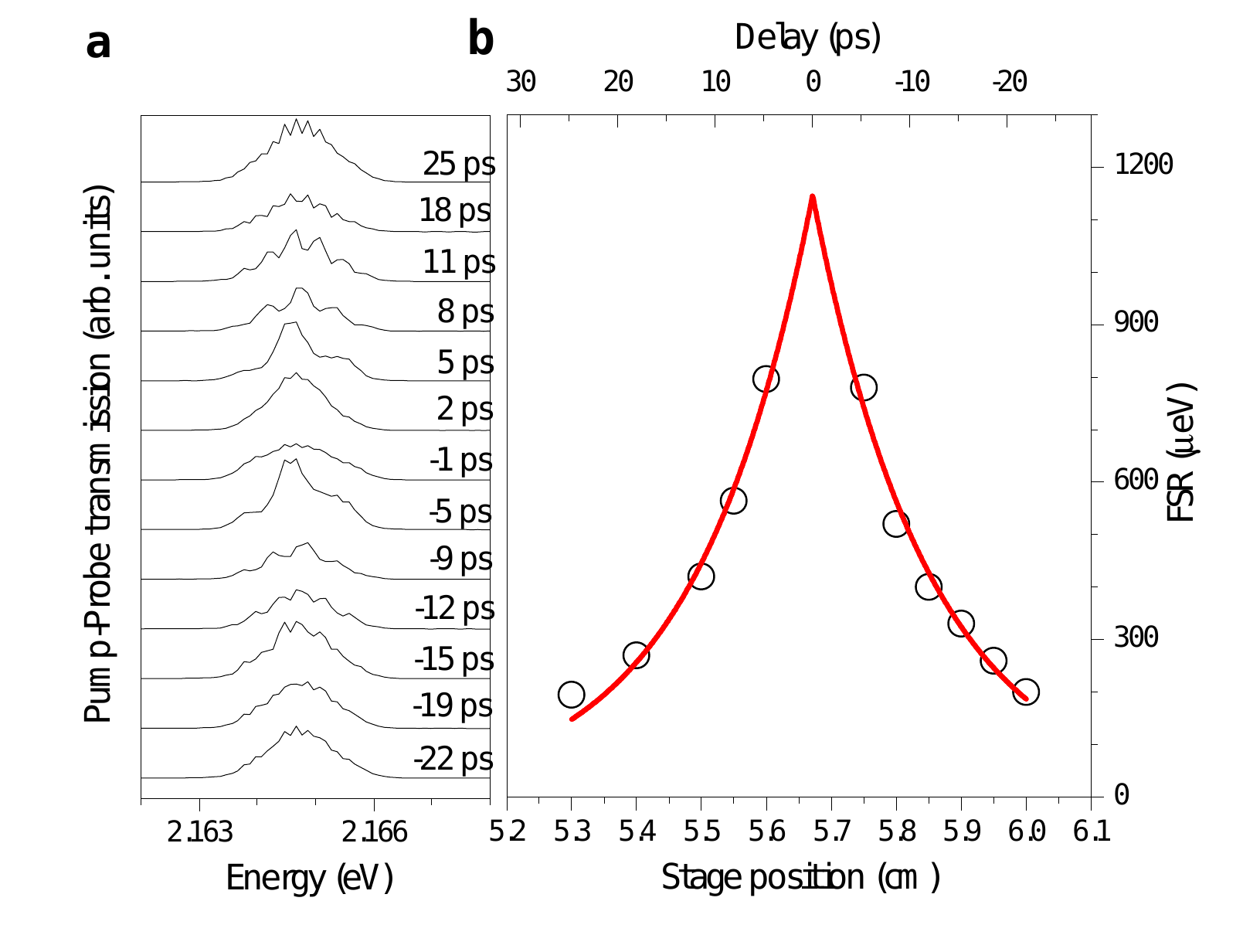}
    \caption{\textcolor{black}{\textbf{Pump-probe signal on substrate without the sample. a} transmitted probe spectra for different delays displays interference with residual pump beam not fully rejected by polarisers. Intensity of probe (pump) beam is 20~$\upmu$J cm$^{-2}$ (20~mJ cm$^{-2}$). Period of modulation in energy or free spectral range (FSR) increases closer to zero delay. \textbf{b} FSR plot (extracted from spectra in \textbf{a} ) as a function of probe delay stage position (bottom axis) and/or time delay (top axis). Time zero between arrival of two pulses is defined from the fit of FSR with exponential function with accuracy of $\pm 0.25$ ps. }}
    \label{Fig:Dzero}
\end{figure}

\textcolor{black}{\section{Pump-probe zero delay point}}
\textcolor{black}{The interference between residual pump and probe pulses results in modulation of the spectra at small delays between pulses (see Fig.~\ref{Fig:Dzero}a). The modulation frequency depends on separation of the pulses whereas its visibility depends on the relative intensities of the two pulses. The analysis of modulation at small delay times allow us to define the zero delay between pulses with accuracy of $\pm0.25$ ps (see Fig.~\ref{Fig:Dzero}). Fig.~\ref{Fig:Dzero}a shows pump-probe transmission spectra on glass substrate without the Cu$_2$O at different delay times where pump signal after rejection with a polariser was $\approx$ 3 times bigger than the probe signal. We extract free spectral range (FSR) of the modulated signal and plot it as a function delay stage position in Fig.~\ref{Fig:Dzero}b. Fitting the FSR data allow us to define the zero delay position for the probe delay stage.}

\textcolor{black}{Although we have used much smaller pump powers in the experiment with Cu$_2$O in Fig. 5 and rejected the unwanted pump signal with polarisers on detection small amount of pump still provides enough modulation for the probe signal to interfere with polariton resonance. So we not plot data points in Fig. 5 of the main text for the range of -30 to 37 ps apart from exact time 0 where the frequency of modulation is bigger than the polariton resonance and it doesn't influence the transmitted probe polariton spectrum. }

\newpage
\newpage

%TC:ignore
\newpage
\bibliography{bibliography}


\begin{thebibliography}{56}
\ifx \bisbn   \undefined \def \bisbn  #1{ISBN #1}\fi
\ifx \binits  \undefined \def \binits#1{#1}\fi
\ifx \bauthor  \undefined \def \bauthor#1{#1}\fi
\ifx \batitle  \undefined \def \batitle#1{#1}\fi
\ifx \bjtitle  \undefined \def \bjtitle#1{#1}\fi
\ifx \bvolume  \undefined \def \bvolume#1{\textbf{#1}}\fi
\ifx \byear  \undefined \def \byear#1{#1}\fi
\ifx \bissue  \undefined \def \bissue#1{#1}\fi
\ifx \bfpage  \undefined \def \bfpage#1{#1}\fi
\ifx \blpage  \undefined \def \blpage #1{#1}\fi
\ifx \burl  \undefined \def \burl#1{\textsf{#1}}\fi
\ifx \doiurl  \undefined \def \doiurl#1{\url{https://doi.org/#1}}\fi
\ifx \betal  \undefined \def \betal{\textit{et al.}}\fi
\ifx \binstitute  \undefined \def \binstitute#1{#1}\fi
\ifx \binstitutionaled  \undefined \def \binstitutionaled#1{#1}\fi
\ifx \bctitle  \undefined \def \bctitle#1{#1}\fi
\ifx \beditor  \undefined \def \beditor#1{#1}\fi
\ifx \bpublisher  \undefined \def \bpublisher#1{#1}\fi
\ifx \bbtitle  \undefined \def \bbtitle#1{#1}\fi
\ifx \bedition  \undefined \def \bedition#1{#1}\fi
\ifx \bseriesno  \undefined \def \bseriesno#1{#1}\fi
\ifx \blocation  \undefined \def \blocation#1{#1}\fi
\ifx \bsertitle  \undefined \def \bsertitle#1{#1}\fi
\ifx \bsnm \undefined \def \bsnm#1{#1}\fi
\ifx \bsuffix \undefined \def \bsuffix#1{#1}\fi
\ifx \bparticle \undefined \def \bparticle#1{#1}\fi
\ifx \barticle \undefined \def \barticle#1{#1}\fi
\bibcommenthead
\ifx \bconfdate \undefined \def \bconfdate #1{#1}\fi
\ifx \botherref \undefined \def \botherref #1{#1}\fi
\ifx \url \undefined \def \url#1{\textsf{#1}}\fi
\ifx \bchapter \undefined \def \bchapter#1{#1}\fi
\ifx \bbook \undefined \def \bbook#1{#1}\fi
\ifx \bcomment \undefined \def \bcomment#1{#1}\fi
\ifx \oauthor \undefined \def \oauthor#1{#1}\fi
\ifx \citeauthoryear \undefined \def \citeauthoryear#1{#1}\fi
\ifx \endbibitem  \undefined \def \endbibitem {}\fi
\ifx \bconflocation  \undefined \def \bconflocation#1{#1}\fi
\ifx \arxivurl  \undefined \def \arxivurl#1{\textsf{#1}}\fi
\csname PreBibitemsHook\endcsname

%%% 1
\bibitem{DingNatPhys2022}
\begin{barticle}
\bauthor{\bsnm{Ding}, \binits{D.-S.}},
\bauthor{\bsnm{Liu}, \binits{Z.-K.}},
\bauthor{\bsnm{Shi}, \binits{B.-S.}},
\bauthor{\bsnm{Guo}, \binits{G.-C.}},
\bauthor{\bsnm{M{\o}lmer}, \binits{K.}},
\bauthor{\bsnm{Adams}, \binits{C.S.}}:
\batitle{Enhanced metrology at the critical point of a many-body rydberg atomic system}.
\bjtitle{Nature Physics}
\bvolume{18}(\bissue{12}),
\bfpage{1447}--\blpage{1452}
(\byear{2022}).
\doiurl{10.1038/s41567-022-01777-8}
\end{barticle}
\endbibitem

%%% 2
\bibitem{Fancher2021}
\begin{barticle}
\bauthor{\bsnm{Fancher}, \binits{C.T.}},
\bauthor{\bsnm{Scherer}, \binits{D.R.}},
\bauthor{\bsnm{John}, \binits{M.C.S.}},
\bauthor{\bsnm{Marlow}, \binits{B.L.S.}}:
\batitle{Rydberg atom electric field sensors for communications and sensing}.
\bjtitle{{IEEE} Transactions on Quantum Engineering}
\bvolume{2},
\bfpage{1}--\blpage{13}
(\byear{2021}).
\doiurl{10.1109/tqe.2021.3065227}
\end{barticle}
\endbibitem

%%% 3
\bibitem{Fancher2022}
\begin{botherref}
\oauthor{\bsnm{Fancher}, \binits{C.T.}},
\oauthor{\bsnm{Nicolich}, \binits{K.L.}},
\oauthor{\bsnm{Backes}, \binits{K.M.}},
\oauthor{\bsnm{Malvania}, \binits{N.}},
\oauthor{\bsnm{Cox}, \binits{K.}},
\oauthor{\bsnm{Meyer}, \binits{D.H.}},
\oauthor{\bsnm{Kunz}, \binits{P.D.}},
\oauthor{\bsnm{Hill}, \binits{J.C.}},
\oauthor{\bsnm{Holland}, \binits{W.}},
\oauthor{\bsnm{Schmittberger~Marlow}, \binits{B.L.}}:
{A self-locking Rydberg atom electric field sensor}.
Applied Physics Letters
\textbf{122}(9)
(2023).
\doiurl{10.1063/5.0137127}
\end{botherref}
\endbibitem

%%% 4
\bibitem{GallgherBook2005}
\begin{bbook}
\bauthor{\bsnm{Gallagher}, \binits{T.F.}}:
\bbtitle{Rydberg Atoms}.
\bpublisher{Cambridge University Press}, \blocation{???}
(\byear{2005})
\end{bbook}
\endbibitem

%%% 5
\bibitem{GrahamPRL2019}
\begin{barticle}
\bauthor{\bsnm{Graham}, \binits{T.M.}},
\bauthor{\bsnm{Kwon}, \binits{M.}},
\bauthor{\bsnm{Grinkemeyer}, \binits{B.}},
\bauthor{\bsnm{Marra}, \binits{Z.}},
\bauthor{\bsnm{Jiang}, \binits{X.}},
\bauthor{\bsnm{Lichtman}, \binits{M.T.}},
\bauthor{\bsnm{Sun}, \binits{Y.}},
\bauthor{\bsnm{Ebert}, \binits{M.}},
\bauthor{\bsnm{Saffman}, \binits{M.}}:
\batitle{Rydberg-mediated entanglement in a two-dimensional neutral atom qubit array}.
\bjtitle{Phys. Rev. Lett.}
\bvolume{123},
\bfpage{230501}
(\byear{2019}).
\doiurl{10.1103/PhysRevLett.123.230501}
\end{barticle}
\endbibitem

%%% 6
\bibitem{Tiarks2019}
\begin{barticle}
\bauthor{\bsnm{Daniel}, \binits{T.}},
\bauthor{\bsnm{Steffen}, \binits{S.-E.}},
\bauthor{\bsnm{Thomas}, \binits{S.}},
\bauthor{\bsnm{Gerhard}, \binits{R.}},
\bauthor{\bsnm{Stephan}, \binits{D.}}:
\batitle{A photon–photon quantum gate based on rydberg interactions}.
\bjtitle{Nature Physics}
\bvolume{15},
\bfpage{124}--\blpage{126}
(\byear{2019}).
\doiurl{10.1038/s41567-018-0313-7}
\end{barticle}
\endbibitem

%%% 7
\bibitem{LukinPRL2001}
\begin{barticle}
\bauthor{\bsnm{Lukin}, \binits{M.D.}},
\bauthor{\bsnm{Fleischhauer}, \binits{M.}},
\bauthor{\bsnm{Cote}, \binits{R.}},
\bauthor{\bsnm{Duan}, \binits{L.M.}},
\bauthor{\bsnm{Jaksch}, \binits{D.}},
\bauthor{\bsnm{Cirac}, \binits{J.I.}},
\bauthor{\bsnm{Zoller}, \binits{P.}}:
\batitle{Dipole blockade and quantum information processing in mesoscopic atomic ensembles}.
\bjtitle{Phys. Rev. Lett.}
\bvolume{87},
\bfpage{037901}
(\byear{2001}).
\doiurl{10.1103/PhysRevLett.87.037901}
\end{barticle}
\endbibitem

%%% 8
\bibitem{Urban2009}
\begin{botherref}
\oauthor{\bsnm{Urban}, \binits{E.}},
\oauthor{\bsnm{Johnson}, \binits{T.A.}},
\oauthor{\bsnm{Henage}, \binits{T.}},
\oauthor{\bsnm{Isenhower}, \binits{L.}},
\oauthor{\bsnm{Yavuz}, \binits{D.D.}},
\oauthor{\bsnm{Walker}, \binits{T.G.}},
\oauthor{\bsnm{Saffman}, \binits{M.}}:
Observation of rydberg blockade between two atoms.
Nature Physics
\textbf{5}(2)
(2009).
\doiurl{10.1038/nphys1178}
\end{botherref}
\endbibitem

%%% 9
\bibitem{Gaetan2009}
\begin{botherref}
\oauthor{\bsnm{Gaëtan}, \binits{A.}},
\oauthor{\bsnm{Miroshnychenko}, \binits{Y.}},
\oauthor{\bsnm{Wilk}, \binits{T.}},
\oauthor{\bsnm{Chotia}, \binits{A.}},
\oauthor{\bsnm{Viteau}, \binits{M.}},
\oauthor{\bsnm{Comparat}, \binits{D.}},
\oauthor{\bsnm{Pillet}, \binits{P.}},
\oauthor{\bsnm{Browaeys}, \binits{A.}},
\oauthor{\bsnm{Grangier}, \binits{P.}}:
Observation of collective excitation of two individual atoms in the rydberg blockade regime.
Nature Physics
\textbf{5}(2)
(2009).
\doiurl{10.1038/nphys1183}
\end{botherref}
\endbibitem

%%% 10
\bibitem{Saffman_RMP2010}
\begin{barticle}
\bauthor{\bsnm{Saffman}, \binits{M.}},
\bauthor{\bsnm{Walker}, \binits{T.G.}},
\bauthor{\bsnm{M\o{}lmer}, \binits{K.}}:
\batitle{Quantum information with rydberg atoms}.
\bjtitle{Rev. Mod. Phys.}
\bvolume{82},
\bfpage{2313}--\blpage{2363}
(\byear{2010}).
\doiurl{10.1103/RevModPhys.82.2313}
\end{barticle}
\endbibitem

%%% 11
\bibitem{Wu_2021}
\begin{barticle}
\bauthor{\bsnm{Wu}, \binits{X.}},
\bauthor{\bsnm{Liang}, \binits{X.}},
\bauthor{\bsnm{Tian}, \binits{Y.}},
\bauthor{\bsnm{Yang}, \binits{F.}},
\bauthor{\bsnm{Chen}, \binits{C.}},
\bauthor{\bsnm{Liu}, \binits{Y.-C.}},
\bauthor{\bsnm{Tey}, \binits{M.K.}},
\bauthor{\bsnm{You}, \binits{L.}}:
\batitle{A concise review of rydberg atom based quantum computation and quantum simulation}.
\bjtitle{Chinese Physics B}
\bvolume{30}(\bissue{2}),
\bfpage{020305}
(\byear{2021}).
\doiurl{10.1088/1674-1056/abd76f}
\end{barticle}
\endbibitem

%%% 12
\bibitem{Morgado2021}
\begin{barticle}
\bauthor{\bsnm{Morgado}, \binits{M.}},
\bauthor{\bsnm{Whitlock}, \binits{S.}}:
\batitle{Quantum simulation and computing with rydberg-interacting qubits}.
\bjtitle{AVS Quantum Science}
\bvolume{3}(\bissue{2}),
\bfpage{023501}
(\byear{2021})
{\href{https://arxiv.org/abs/https://doi.org/10.1116/5.0036562}{{https://doi.org/10.1116/5.0036562}}}.
\doiurl{10.1116/5.0036562}
\end{barticle}
\endbibitem

%%% 13
\bibitem{Shi2022}
\begin{barticle}
\bauthor{\bsnm{Shi}, \binits{X.-F.}}:
\batitle{Quantum logic and entanglement by neutral rydberg atoms: methods and fidelity}.
\bjtitle{Quantum Science and Technology}
\bvolume{7}(\bissue{2}),
\bfpage{023002}
(\byear{2022}).
\doiurl{10.1088/2058-9565/ac18b8}
\end{barticle}
\endbibitem

%%% 14
\bibitem{MurrayAdv2016}
\begin{botherref}
\oauthor{\bsnm{Murray}, \binits{C.}},
\oauthor{\bsnm{Pohl}, \binits{T.}}:
Chapter seven - quantum and nonlinear optics in strongly interacting atomic ensembles.
Advances In Atomic, Molecular, and Optical Physics,
vol. 65,
pp. 321--372.
Academic Press
(2016).
\doiurl{10.1016/bs.aamop.2016.04.005}.
\url{https://www.sciencedirect.com/science/article/pii/S1049250X1630009X}
\end{botherref}
\endbibitem

%%% 15
\bibitem{Fleischhauer2005}
\begin{barticle}
\bauthor{\bsnm{Fleischhauer}, \binits{M.}},
\bauthor{\bsnm{Imamoglu}, \binits{A.}},
\bauthor{\bsnm{Marangos}, \binits{J.P.}}:
\batitle{Electromagnetically induced transparency: Optics in coherent media}.
\bjtitle{Rev. Mod. Phys.}
\bvolume{77},
\bfpage{633}--\blpage{673}
(\byear{2005}).
\doiurl{10.1103/RevModPhys.77.633}
\end{barticle}
\endbibitem

%%% 16
\bibitem{Firstenberg_2016}
\begin{barticle}
\bauthor{\bsnm{Firstenberg}, \binits{O.}},
\bauthor{\bsnm{Adams}, \binits{C.S.}},
\bauthor{\bsnm{Hofferberth}, \binits{S.}}:
\batitle{Nonlinear quantum optics mediated by rydberg interactions}.
\bjtitle{Journal of Physics B: Atomic, Molecular and Optical Physics}
\bvolume{49}(\bissue{15}),
\bfpage{152003}
(\byear{2016}).
\doiurl{10.1088/0953-4075/49/15/152003}
\end{barticle}
\endbibitem

%%% 17
\bibitem{Gu2021}
\begin{botherref}
\oauthor{\bsnm{Gu}, \binits{J.}},
\oauthor{\bsnm{Walther}, \binits{V.}},
\oauthor{\bsnm{Waldecker}, \binits{L.}},
\oauthor{\bsnm{Rhodes}, \binits{D.}},
\oauthor{\bsnm{Raja}, \binits{A.}},
\oauthor{\bsnm{Hone}, \binits{J.C.}},
\oauthor{\bsnm{Heinz}, \binits{T.F.}},
\oauthor{\bsnm{K{\'{e}}na-Cohen}, \binits{S.}},
\oauthor{\bsnm{Pohl}, \binits{T.}},
\oauthor{\bsnm{Menon}, \binits{V.M.}}:
Enhanced nonlinear interaction of polaritons via excitonic rydberg states in monolayer {WSe}2.
Nature Communications
\textbf{12}(1)
(2021).
\doiurl{10.1038/s41467-021-22537-x}
\end{botherref}
\endbibitem

%%% 18
\bibitem{Bao2019}
\begin{barticle}
\bauthor{\bsnm{Bao}, \binits{W.}},
\bauthor{\bsnm{Liu}, \binits{X.}},
\bauthor{\bsnm{Xue}, \binits{F.}},
\bauthor{\bsnm{Zheng}, \binits{F.}},
\bauthor{\bsnm{Tao}, \binits{R.}},
\bauthor{\bsnm{Wang}, \binits{S.}},
\bauthor{\bsnm{Xia}, \binits{Y.}},
\bauthor{\bsnm{Zhao}, \binits{M.}},
\bauthor{\bsnm{Kim}, \binits{J.}},
\bauthor{\bsnm{Yang}, \binits{S.}},
\bauthor{\bsnm{Li}, \binits{Q.}},
\bauthor{\bsnm{Wang}, \binits{Y.}},
\bauthor{\bsnm{Wang}, \binits{Y.}},
\bauthor{\bsnm{Wang}, \binits{L.-W.}},
\bauthor{\bsnm{MacDonald}, \binits{A.H.}},
\bauthor{\bsnm{Zhang}, \binits{X.}}:
\batitle{Observation of rydberg exciton polaritons and their condensate in a perovskite cavity}.
\bjtitle{Proceedings of the National Academy of Sciences}
\bvolume{116}(\bissue{41}),
\bfpage{20274}--\blpage{20279}
(\byear{2019}).
\doiurl{10.1073/pnas.1909948116}
\end{barticle}
\endbibitem

%%% 19
\bibitem{Kazimierczuk2014}
\begin{barticle}
\bauthor{\bsnm{Kazimierczuk}, \binits{T.}},
\bauthor{\bsnm{Fr\"{o}hlich}, \binits{D.}},
\bauthor{\bsnm{Scheel}, \binits{S.}},
\bauthor{\bsnm{Stolz}, \binits{H.}},
\bauthor{\bsnm{Bayer}, \binits{M.}}:
\batitle{Giant rydberg excitons in the copper oxide \cutwoo}.
\bjtitle{Nature}
\bvolume{514}(\bissue{7522}),
\bfpage{343}--\blpage{347}
(\byear{2014}).
\doiurl{10.1038/nature13832}
\end{barticle}
\endbibitem

%%% 20
\bibitem{Amann2020}
\begin{barticle}
\bauthor{\bsnm{A{\ss}mann}, \binits{M.}},
\bauthor{\bsnm{Bayer}, \binits{M.}}:
\batitle{Semiconductor rydberg physics}.
\bjtitle{Advanced Quantum Technologies}
\bvolume{3}(\bissue{11}),
\bfpage{1900134}
(\byear{2020}).
\doiurl{10.1002/qute.201900134}
\end{barticle}
\endbibitem

%%% 21
\bibitem{Morin2022}
\begin{botherref}
\oauthor{\bsnm{Morin}, \binits{C.}},
\oauthor{\bsnm{Tignon}, \binits{J.}},
\oauthor{\bsnm{Mangeney}, \binits{J.}},
\oauthor{\bsnm{Dhillon}, \binits{S.}},
\oauthor{\bsnm{Czajkowski}, \binits{G.}},
\oauthor{\bsnm{Karpi{\'{n}}ski}, \binits{K.}},
\oauthor{\bsnm{Zieli{\'{n}}ska-Raczy{\'{n}}ska}, \binits{S.}},
\oauthor{\bsnm{Ziemkiewicz}, \binits{D.}},
\oauthor{\bsnm{Boulier}, \binits{T.}}:
Self-kerr effect across the yellow rydberg series of excitons in \cutwoo.
Physical Review Letters
\textbf{129}(13)
(2022).
\doiurl{10.1103/physrevlett.129.137401}
\end{botherref}
\endbibitem

%%% 22
\bibitem{Steinhauer2020}
\begin{barticle}
\bauthor{\bsnm{Steinhauer}, \binits{S.}},
\bauthor{\bsnm{Versteegh}, \binits{M.A.M.}},
\bauthor{\bsnm{Gyger}, \binits{S.}},
\bauthor{\bsnm{Elshaari}, \binits{A.W.}},
\bauthor{\bsnm{Kunert}, \binits{B.}},
\bauthor{\bsnm{Mysyrowicz}, \binits{A.}},
\bauthor{\bsnm{Zwiller}, \binits{V.}}:
\batitle{Rydberg excitons in \uppercase{C}u$_{2}$\uppercase{O} microcrystals grown on a silicon platform}.
\bjtitle{Communications Materials}
\bvolume{1}(\bissue{1}),
\bfpage{11}
(\byear{2020}).
\doiurl{10.1038/s43246-020-0013-6}
\end{barticle}
\endbibitem

%%% 23
\bibitem{Versteegh2021}
\begin{barticle}
\bauthor{\bsnm{Versteegh}, \binits{M.A.M.}},
\bauthor{\bsnm{Steinhauer}, \binits{S.}},
\bauthor{\bsnm{Bajo}, \binits{J.}},
\bauthor{\bsnm{Lettner}, \binits{T.}},
\bauthor{\bsnm{Soro}, \binits{A.}},
\bauthor{\bsnm{Romanova}, \binits{A.}},
\bauthor{\bsnm{Gyger}, \binits{S.}},
\bauthor{\bsnm{Schweickert}, \binits{L.}},
\bauthor{\bsnm{Mysyrowicz}, \binits{A.}},
\bauthor{\bsnm{Zwiller}, \binits{V.}}:
\batitle{Giant rydberg excitons in \uppercase{C}u$_{2}$\uppercase{O} probed by photoluminescence excitation spectroscopy}.
\bjtitle{Phys. Rev. B}
\bvolume{104},
\bfpage{245206}
(\byear{2021}).
\doiurl{10.1103/PhysRevB.104.245206}
\end{barticle}
\endbibitem

%%% 24
\bibitem{WaltherPRL2020}
\begin{barticle}
\bauthor{\bsnm{Walther}, \binits{V.}},
\bauthor{\bsnm{Pohl}, \binits{T.}}:
\batitle{Plasma-enhanced interaction and optical nonlinearities of ${\mathrm{cu}}_{2}\mathrm{O}$ rydberg excitons}.
\bjtitle{Phys. Rev. Lett.}
\bvolume{125},
\bfpage{097401}
(\byear{2020}).
\doiurl{10.1103/PhysRevLett.125.097401}
\end{barticle}
\endbibitem

%%% 25
\bibitem{Hecktter2021}
\begin{botherref}
\oauthor{\bsnm{Heck\"{o}tter}, \binits{J.}},
\oauthor{\bsnm{Walther}, \binits{V.}},
\oauthor{\bsnm{Scheel}, \binits{S.}},
\oauthor{\bsnm{Bayer}, \binits{M.}},
\oauthor{\bsnm{Pohl}, \binits{T.}},
\oauthor{\bsnm{A{\ss}mann}, \binits{M.}}:
Asymmetric rydberg blockade of giant excitons in cuprous oxide.
Nature Communications
\textbf{12}(1)
(2021).
\doiurl{10.1038/s41467-021-23852-z}
\end{botherref}
\endbibitem

%%% 26
\bibitem{CarusottoCiuti2013}
\begin{barticle}
\bauthor{\bsnm{Carusotto}, \binits{I.}},
\bauthor{\bsnm{Ciuti}, \binits{C.}}:
\batitle{Quantum fluids of light}.
\bjtitle{Rev. Mod. Phys.}
\bvolume{85},
\bfpage{299}--\blpage{366}
(\byear{2013}).
\doiurl{10.1103/RevModPhys.85.299}
\end{barticle}
\endbibitem

%%% 27
\bibitem{Byrnes2014}
\begin{barticle}
\bauthor{\bsnm{Byrnes}, \binits{T.}},
\bauthor{\bsnm{Kim}, \binits{N.Y.}},
\bauthor{\bsnm{Yamamoto}, \binits{Y.}}:
\batitle{Exciton{\textendash}polariton condensates}.
\bjtitle{Nature Physics}
\bvolume{10}(\bissue{11}),
\bfpage{803}--\blpage{813}
(\byear{2014}).
\doiurl{10.1038/nphys3143}
\end{barticle}
\endbibitem

%%% 28
\bibitem{Kuriakose2022}
\begin{barticle}
\bauthor{\bsnm{Kuriakose}, \binits{T.}},
\bauthor{\bsnm{Walker}, \binits{P.M.}},
\bauthor{\bsnm{Dowling}, \binits{T.}},
\bauthor{\bsnm{Kyriienko}, \binits{O.}},
\bauthor{\bsnm{Shelykh}, \binits{I.A.}},
\bauthor{\bsnm{St-Jean}, \binits{P.}},
\bauthor{\bsnm{Zambon}, \binits{N.C.}},
\bauthor{\bsnm{Lema{\^{\i}}tre}, \binits{A.}},
\bauthor{\bsnm{Sagnes}, \binits{I.}},
\bauthor{\bsnm{Legratiet}, \binits{L.}},
\bauthor{\bsnm{Harouri}, \binits{A.}},
\bauthor{\bsnm{Ravets}, \binits{S.}},
\bauthor{\bsnm{Skolnick}, \binits{M.S.}},
\bauthor{\bsnm{Amo}, \binits{A.}},
\bauthor{\bsnm{Bloch}, \binits{J.}},
\bauthor{\bsnm{Krizhanovskii}, \binits{D.N.}}:
\batitle{Few-photon all-optical phase rotation in a quantum-well micropillar cavity}.
\bjtitle{Nature Photonics}
\bvolume{16}(\bissue{8}),
\bfpage{566}--\blpage{569}
(\byear{2022}).
\doiurl{10.1038/s41566-022-01019-6}
\end{barticle}
\endbibitem

%%% 29
\bibitem{Delteil2019}
\begin{barticle}
\bauthor{\bsnm{Delteil}, \binits{A.}},
\bauthor{\bsnm{Fink}, \binits{T.}},
\bauthor{\bsnm{Schade}, \binits{A.}},
\bauthor{\bsnm{H\"{o}fling}, \binits{S.}},
\bauthor{\bsnm{Schneider}, \binits{C.}},
\bauthor{\bsnm{{\.{I}}mamo{\u{g}}lu}, \binits{A.}}:
\batitle{Towards polariton blockade of confined exciton{\textendash}polaritons}.
\bjtitle{Nature Materials}
\bvolume{18}(\bissue{3}),
\bfpage{219}--\blpage{222}
(\byear{2019}).
\doiurl{10.1038/s41563-019-0282-y}
\end{barticle}
\endbibitem

%%% 30
\bibitem{MuozMatutano2019}
\begin{barticle}
\bauthor{\bsnm{Mu{\~{n}}oz-Matutano}, \binits{G.}},
\bauthor{\bsnm{Wood}, \binits{A.}},
\bauthor{\bsnm{Johnsson}, \binits{M.}},
\bauthor{\bsnm{Vidal}, \binits{X.}},
\bauthor{\bsnm{Baragiola}, \binits{B.Q.}},
\bauthor{\bsnm{Reinhard}, \binits{A.}},
\bauthor{\bsnm{Lema{\^{\i}}tre}, \binits{A.}},
\bauthor{\bsnm{Bloch}, \binits{J.}},
\bauthor{\bsnm{Amo}, \binits{A.}},
\bauthor{\bsnm{Nogues}, \binits{G.}},
\bauthor{\bsnm{Besga}, \binits{B.}},
\bauthor{\bsnm{Richard}, \binits{M.}},
\bauthor{\bsnm{Volz}, \binits{T.}}:
\batitle{Emergence of quantum correlations from interacting fibre-cavity polaritons}.
\bjtitle{Nature Materials}
\bvolume{18}(\bissue{3}),
\bfpage{213}--\blpage{218}
(\byear{2019}).
\doiurl{10.1038/s41563-019-0281-z}
\end{barticle}
\endbibitem

%%% 31
\bibitem{Chernikov2014}
\begin{barticle}
\bauthor{\bsnm{Chernikov}, \binits{A.}},
\bauthor{\bsnm{Berkelbach}, \binits{T.C.}},
\bauthor{\bsnm{Hill}, \binits{H.M.}},
\bauthor{\bsnm{Rigosi}, \binits{A.}},
\bauthor{\bsnm{Li}, \binits{Y.}},
\bauthor{\bsnm{Aslan}, \binits{B.}},
\bauthor{\bsnm{Reichman}, \binits{D.R.}},
\bauthor{\bsnm{Hybertsen}, \binits{M.S.}},
\bauthor{\bsnm{Heinz}, \binits{T.F.}}:
\batitle{Exciton binding energy and nonhydrogenic rydberg series in monolayer ${\mathrm{ws}}_{2}$}.
\bjtitle{Phys. Rev. Lett.}
\bvolume{113},
\bfpage{076802}
(\byear{2014}).
\doiurl{10.1103/PhysRevLett.113.076802}
\end{barticle}
\endbibitem

%%% 32
\bibitem{Coriolano2022}
\begin{botherref}
\oauthor{\bsnm{Coriolano}, \binits{A.}},
\oauthor{\bsnm{Polimeno}, \binits{L.}},
\oauthor{\bsnm{Pugliese}, \binits{M.}},
\oauthor{\bsnm{Cannavale}, \binits{A.}},
\oauthor{\bsnm{Trypogeorgos}, \binits{D.}},
\oauthor{\bsnm{Renzo}, \binits{A.D.}},
\oauthor{\bsnm{Ardizzone}, \binits{V.}},
\oauthor{\bsnm{Rizzo}, \binits{A.}},
\oauthor{\bsnm{Ballarini}, \binits{D.}},
\oauthor{\bsnm{Gigli}, \binits{G.}},
\oauthor{\bsnm{Maiorano}, \binits{V.}},
\oauthor{\bsnm{Rosyadi}, \binits{A.S.}},
\oauthor{\bsnm{Chuang}, \binits{C.-A.}},
\oauthor{\bsnm{Ho}, \binits{C.-H.}},
\oauthor{\bsnm{Marco}, \binits{L.D.}},
\oauthor{\bsnm{Sanvitto}, \binits{D.}},
\oauthor{\bsnm{Giorgi}, \binits{M.D.}}:
Rydberg polaritons in $\mathrm{ReS}_2$ crystals.
Science Advances
\textbf{8}(47)
(2022).
\doiurl{10.1126/sciadv.add8857}
\end{botherref}
\endbibitem

%%% 33
\bibitem{Orfanakis2022}
\begin{barticle}
\bauthor{\bsnm{Orfanakis}, \binits{K.}},
\bauthor{\bsnm{Rajendran}, \binits{S.K.}},
\bauthor{\bsnm{Walther}, \binits{V.}},
\bauthor{\bsnm{Volz}, \binits{T.}},
\bauthor{\bsnm{Pohl}, \binits{T.}},
\bauthor{\bsnm{Ohadi}, \binits{H.}}:
\batitle{Rydberg exciton{\textendash}polaritons in a \cutwoo microcavity}.
\bjtitle{Nature Materials}
\bvolume{21}(\bissue{7}),
\bfpage{767}--\blpage{772}
(\byear{2022}).
\doiurl{10.1038/s41563-022-01230-4}
\end{barticle}
\endbibitem

%%% 34
\bibitem{Fieramosca_2019}
\begin{botherref}
\oauthor{\bsnm{Fieramosca}, \binits{A.}},
\oauthor{\bsnm{Polimeno}, \binits{L.}},
\oauthor{\bsnm{Ardizzone}, \binits{V.}},
\oauthor{\bsnm{Marco}, \binits{L.D.}},
\oauthor{\bsnm{Pugliese}, \binits{M.}},
\oauthor{\bsnm{Maiorano}, \binits{V.}},
\oauthor{\bsnm{Giorgi}, \binits{M.D.}},
\oauthor{\bsnm{Dominici}, \binits{L.}},
\oauthor{\bsnm{Gigli}, \binits{G.}},
\oauthor{\bsnm{Gerace}, \binits{D.}},
\oauthor{\bsnm{Ballarini}, \binits{D.}},
\oauthor{\bsnm{Sanvitto}, \binits{D.}}:
Two-dimensional hybrid perovskites sustaining strong polariton interactions at room temperature.
Science Advances
\textbf{5}(5)
(2019).
\doiurl{10.1126/sciadv.aav9967}
\end{botherref}
\endbibitem

%%% 35
\bibitem{HeckotterPRB2017}
\begin{barticle}
\bauthor{\bsnm{Heck\"otter}, \binits{J.}},
\bauthor{\bsnm{Freitag}, \binits{M.}},
\bauthor{\bsnm{Fr\"ohlich}, \binits{D.}},
\bauthor{\bsnm{A\ss{}mann}, \binits{M.}},
\bauthor{\bsnm{Bayer}, \binits{M.}},
\bauthor{\bsnm{Semina}, \binits{M.A.}},
\bauthor{\bsnm{Glazov}, \binits{M.M.}}:
\batitle{Scaling laws of rydberg excitons}.
\bjtitle{Phys. Rev. B}
\bvolume{96},
\bfpage{125142}
(\byear{2017}).
\doiurl{10.1103/PhysRevB.96.125142}
\end{barticle}
\endbibitem

%%% 36
\bibitem{Emmanuele2020}
\begin{barticle}
\bauthor{\bsnm{Emmanuele}, \binits{R.P.A.}},
\bauthor{\bsnm{Sich}, \binits{M.}},
\bauthor{\bsnm{Kyriienko}, \binits{O.}},
\bauthor{\bsnm{Shahnazaryan}, \binits{V.}},
\bauthor{\bsnm{Withers}, \binits{F.}},
\bauthor{\bsnm{Catanzaro}, \binits{A.}},
\bauthor{\bsnm{Walker}, \binits{P.M.}},
\bauthor{\bsnm{Benimetskiy}, \binits{F.A.}},
\bauthor{\bsnm{Skolnick}, \binits{M.S.}},
\bauthor{\bsnm{I.}, \binits{T.A.}},
\bauthor{\bsnm{Shelykh}, \binits{I.A.}},
\bauthor{\bsnm{Krizhanovskii}, \binits{D.N.}}:
\batitle{Highly nonlinear trion-polaritons in a monolayer semiconductor}.
\bjtitle{Nature Communications}
\bvolume{11},
\bfpage{3589}
(\byear{2020}).
\doiurl{10.1038/s41467-020-17340-z}
\end{barticle}
\endbibitem

%%% 37
\bibitem{Brichkin2011}
\begin{barticle}
\bauthor{\bsnm{Brichkin}, \binits{A.S.}},
\bauthor{\bsnm{Novikov}, \binits{S.I.}},
\bauthor{\bsnm{Larionov}, \binits{A.V.}},
\bauthor{\bsnm{Kulakovskii}, \binits{V.D.}},
\bauthor{\bsnm{Glazov}, \binits{M.M.}},
\bauthor{\bsnm{Schneider}, \binits{C.}},
\bauthor{\bsnm{H\"ofling}, \binits{S.}},
\bauthor{\bsnm{Kamp}, \binits{M.}},
\bauthor{\bsnm{Forchel}, \binits{A.}}:
\batitle{Effect of coulomb interaction on exciton-polariton condensates in gaas pillar microcavities}.
\bjtitle{Phys. Rev. B}
\bvolume{84},
\bfpage{195301}
(\byear{2011}).
\doiurl{10.1103/PhysRevB.84.195301}
\end{barticle}
\endbibitem

%%% 38
\bibitem{Zhang2022}
\begin{barticle}
\bauthor{\bsnm{Zhang}, \binits{L.}},
\bauthor{\bsnm{Walther}, \binits{V.}},
\bauthor{\bsnm{M{\o{}}lmer}, \binits{K.}},
\bauthor{\bsnm{Pohl}, \binits{T.}}:
\batitle{Photon-photon interactions in {R}ydberg-atom arrays}.
\bjtitle{{Quantum}}
\bvolume{6},
\bfpage{674}
(\byear{2022}).
\doiurl{10.22331/q-2022-03-30-674}
\end{barticle}
\endbibitem

%%% 39
\bibitem{Khitrova:RMP71(1999)}
\begin{barticle}
\bauthor{\bsnm{Khitrova}, \binits{G.}},
\bauthor{\bsnm{Gibbs}, \binits{H.M.}},
\bauthor{\bsnm{Jahnke}, \binits{F.}},
\bauthor{\bsnm{Kira}, \binits{M.}},
\bauthor{\bsnm{Koch}, \binits{S.W.}}:
\batitle{Nonlinear optics of normal-mode-coupling semiconductor microcavities}.
\bjtitle{Rev. Mod. Phys.}
\bvolume{71},
\bfpage{1591}--\blpage{1639}
(\byear{1999}).
\doiurl{10.1103/RevModPhys.71.1591}
\end{barticle}
\endbibitem

%%% 40
\bibitem{Walker2015}
\begin{botherref}
\oauthor{\bsnm{Walker}, \binits{P.M.}},
\oauthor{\bsnm{Tinkler}, \binits{L.}},
\oauthor{\bsnm{Skryabin}, \binits{D.V.}},
\oauthor{\bsnm{Yulin}, \binits{A.}},
\oauthor{\bsnm{Royall}, \binits{B.}},
\oauthor{\bsnm{Farrer}, \binits{I.}},
\oauthor{\bsnm{Ritchie}, \binits{D.A.}},
\oauthor{\bsnm{Skolnick}, \binits{M.S.}},
\oauthor{\bsnm{Krizhanovskii}, \binits{D.N.}}:
Ultra-low-power hybrid light–matter solitons.
Nature Communications
\textbf{6}
(2015).
\doiurl{10.1038/ncomms9317}
\end{botherref}
\endbibitem

%%% 41
\bibitem{Walther:PRB98(2018)}
\begin{barticle}
\bauthor{\bsnm{Walther}, \binits{V.}},
\bauthor{\bsnm{Kr\"uger}, \binits{S.O.}},
\bauthor{\bsnm{Scheel}, \binits{S.}},
\bauthor{\bsnm{Pohl}, \binits{T.}}:
\batitle{Interactions between rydberg excitons in \uppercase{C}u$_{2}$\uppercase{O}}.
\bjtitle{Phys. Rev. B}
\bvolume{98},
\bfpage{165201}
(\byear{2018}).
\doiurl{10.1103/PhysRevB.98.165201}
\end{barticle}
\endbibitem

%%% 42
\bibitem{PWalker2017}
\begin{barticle}
\bauthor{\bsnm{Walker}, \binits{P.M.}},
\bauthor{\bsnm{Tinkler}, \binits{L.}},
\bauthor{\bsnm{Royall}, \binits{B.}},
\bauthor{\bsnm{Skryabin}, \binits{D.V.}},
\bauthor{\bsnm{Farrer}, \binits{I.}},
\bauthor{\bsnm{Ritchie}, \binits{D.A.}},
\bauthor{\bsnm{Skolnick}, \binits{M.S.}},
\bauthor{\bsnm{Krizhanovskii}, \binits{D.N.}}:
\batitle{Dark solitons in high velocity waveguide polariton fluids}.
\bjtitle{Phys. Rev. Lett.}
\bvolume{119},
\bfpage{097403}
(\byear{2017}).
\doiurl{10.1103/PhysRevLett.119.097403}
\end{barticle}
\endbibitem

%%% 43
\bibitem{MysyPRL1979}
\begin{barticle}
\bauthor{\bsnm{Mysyrowicz}, \binits{A.}},
\bauthor{\bsnm{Hulin}, \binits{D.}},
\bauthor{\bsnm{Antonetti}, \binits{A.}}:
\batitle{Long exciton lifetime in ${\mathrm{cu}}_{2}$o}.
\bjtitle{Phys. Rev. Lett.}
\bvolume{43},
\bfpage{1123}--\blpage{1126}
(\byear{1979}).
\doiurl{10.1103/PhysRevLett.43.1123}
\end{barticle}
\endbibitem

%%% 44
\bibitem{Rogers2022}
\begin{barticle}
\bauthor{\bsnm{Rogers}, \binits{J.P.}},
\bauthor{\bsnm{Gallagher}, \binits{L.A.P.}},
\bauthor{\bsnm{Pizzey}, \binits{D.}},
\bauthor{\bsnm{Pritchett}, \binits{J.D.}},
\bauthor{\bsnm{Adams}, \binits{C.S.}},
\bauthor{\bsnm{Jones}, \binits{M.P.A.}},
\bauthor{\bsnm{Hodges}, \binits{C.}},
\bauthor{\bsnm{Langbein}, \binits{W.}},
\bauthor{\bsnm{Lynch}, \binits{S.A.}}:
\batitle{High-resolution nanosecond spectroscopy of even-parity rydberg excitons in ${\mathrm{cu}}_{2}\mathrm{O}$}.
\bjtitle{Phys. Rev. B}
\bvolume{105},
\bfpage{115206}
(\byear{2022}).
\doiurl{10.1103/PhysRevB.105.115206}
\end{barticle}
\endbibitem

%%% 45
\bibitem{HeckotterPRL2018}
\begin{barticle}
\bauthor{\bsnm{Heck\"otter}, \binits{J.}},
\bauthor{\bsnm{Freitag}, \binits{M.}},
\bauthor{\bsnm{Fr\"ohlich}, \binits{D.}},
\bauthor{\bsnm{A\ss{}mann}, \binits{M.}},
\bauthor{\bsnm{Bayer}, \binits{M.}},
\bauthor{\bsnm{Gr\"unwald}, \binits{P.}},
\bauthor{\bsnm{Sch\"one}, \binits{F.}},
\bauthor{\bsnm{Semkat}, \binits{D.}},
\bauthor{\bsnm{Stolz}, \binits{H.}},
\bauthor{\bsnm{Scheel}, \binits{S.}}:
\batitle{Rydberg excitons in the presence of an ultralow-density electron-hole plasma}.
\bjtitle{Phys. Rev. Lett.}
\bvolume{121},
\bfpage{097401}
(\byear{2018}).
\doiurl{10.1103/PhysRevLett.121.097401}
\end{barticle}
\endbibitem

%%% 46
\bibitem{StolzPRB2022}
\begin{barticle}
\bauthor{\bsnm{Stolz}, \binits{H.}},
\bauthor{\bsnm{Semkat}, \binits{D.}},
\bauthor{\bsnm{Schwartz}, \binits{R.}},
\bauthor{\bsnm{Heck\"otter}, \binits{J.}},
\bauthor{\bsnm{A\ss{}mann}, \binits{M.}},
\bauthor{\bsnm{Kraeft}, \binits{W.-D.}},
\bauthor{\bsnm{Fehske}, \binits{H.}},
\bauthor{\bsnm{Bayer}, \binits{M.}}:
\batitle{Scrutinizing the debye plasma model: Rydberg excitons unravel the properties of low-density plasmas in semiconductors}.
\bjtitle{Phys. Rev. B}
\bvolume{105},
\bfpage{075204}
(\byear{2022}).
\doiurl{10.1103/PhysRevB.105.075204}
\end{barticle}
\endbibitem

%%% 47
\bibitem{Walther2018}
\begin{botherref}
\oauthor{\bsnm{Walther}, \binits{V.}},
\oauthor{\bsnm{Johne}, \binits{R.}},
\oauthor{\bsnm{Pohl}, \binits{T.}}:
Giant optical nonlinearities from rydberg excitons in semiconductor microcavities.
Nature Communications
\textbf{9}(1)
(2018).
\doiurl{10.1038/s41467-018-03742-7}
\end{botherref}
\endbibitem

%%% 48
\bibitem{PhysRevMaterials.5.084602}
\begin{barticle}
\bauthor{\bsnm{Lynch}, \binits{S.A.}},
\bauthor{\bsnm{Hodges}, \binits{C.}},
\bauthor{\bsnm{Mandal}, \binits{S.}},
\bauthor{\bsnm{Langbein}, \binits{W.}},
\bauthor{\bsnm{Singh}, \binits{R.P.}},
\bauthor{\bsnm{Gallagher}, \binits{L.A.P.}},
\bauthor{\bsnm{Pritchett}, \binits{J.D.}},
\bauthor{\bsnm{Pizzey}, \binits{D.}},
\bauthor{\bsnm{Rogers}, \binits{J.P.}},
\bauthor{\bsnm{Adams}, \binits{C.S.}},
\bauthor{\bsnm{Jones}, \binits{M.P.A.}}:
\batitle{Rydberg excitons in synthetic cuprous oxide ${\mathrm{cu}}_{2}\mathrm{O}$}.
\bjtitle{Phys. Rev. Materials}
\bvolume{5},
\bfpage{084602}
(\byear{2021}).
\doiurl{10.1103/PhysRevMaterials.5.084602}
\end{barticle}
\endbibitem

%%% 49
\bibitem{Ferrera2019}
\begin{barticle}
\bauthor{\bsnm{Ferrera}, \binits{M.}},
\bauthor{\bsnm{Magnozzi}, \binits{M.}},
\bauthor{\bsnm{Bisio}, \binits{F.}},
\bauthor{\bsnm{Canepa}, \binits{M.}}:
\batitle{Temperature-dependent permittivity of silver and implications for thermoplasmonics}.
\bjtitle{Phys. Rev. Mater.}
\bvolume{3},
\bfpage{105201}
(\byear{2019}).
\doiurl{10.1103/PhysRevMaterials.3.105201}
\end{barticle}
\endbibitem

%%% 50
\bibitem{Carusotto_2010}
\begin{barticle}
\bauthor{\bsnm{Carusotto}, \binits{I.}},
\bauthor{\bsnm{Volz}, \binits{T.}},
\bauthor{\bsnm{Imamo{\u{g}}lu}, \binits{A.}}:
\batitle{Feshbach blockade: Single-photon nonlinear optics using resonantly enhanced cavity polariton scattering from biexciton states}.
\bjtitle{{EPL} (Europhysics Letters)}
\bvolume{90}(\bissue{3}),
\bfpage{37001}
(\byear{2010}).
\doiurl{10.1209/0295-5075/90/37001}
\end{barticle}
\endbibitem

%%% 51
\bibitem{Ferr2011}
\begin{barticle}
\bauthor{\bsnm{Ferrier}, \binits{L.}},
\bauthor{\bsnm{Wertz}, \binits{E.}},
\bauthor{\bsnm{Johne}, \binits{R.}},
\bauthor{\bsnm{Solnyshkov}, \binits{D.D.}},
\bauthor{\bsnm{Senellart}, \binits{P.}},
\bauthor{\bsnm{Sagnes}, \binits{I.}},
\bauthor{\bsnm{Lema\^{\i}tre}, \binits{A.}},
\bauthor{\bsnm{Malpuech}, \binits{G.}},
\bauthor{\bsnm{Bloch}, \binits{J.}}:
\batitle{Interactions in confined polariton condensates}.
\bjtitle{Phys. Rev. Lett.}
\bvolume{106},
\bfpage{126401}
(\byear{2011}).
\doiurl{10.1103/PhysRevLett.106.126401}
\end{barticle}
\endbibitem

%%% 52
\bibitem{EstrPRB2019}
\begin{barticle}
\bauthor{\bsnm{Estrecho}, \binits{E.}},
\bauthor{\bsnm{Gao}, \binits{T.}},
\bauthor{\bsnm{Bobrovska}, \binits{N.}},
\bauthor{\bsnm{Comber-Todd}, \binits{D.}},
\bauthor{\bsnm{Fraser}, \binits{M.D.}},
\bauthor{\bsnm{Steger}, \binits{M.}},
\bauthor{\bsnm{West}, \binits{K.}},
\bauthor{\bsnm{Pfeiffer}, \binits{L.N.}},
\bauthor{\bsnm{Levinsen}, \binits{J.}},
\bauthor{\bsnm{Parish}, \binits{M.M.}},
\bauthor{\bsnm{Liew}, \binits{T.C.H.}},
\bauthor{\bsnm{Matuszewski}, \binits{M.}},
\bauthor{\bsnm{Snoke}, \binits{D.W.}},
\bauthor{\bsnm{Truscott}, \binits{A.G.}},
\bauthor{\bsnm{Ostrovskaya}, \binits{E.A.}}:
\batitle{Direct measurement of polariton-polariton interaction strength in the thomas-fermi regime of exciton-polariton condensation}.
\bjtitle{Phys. Rev. B}
\bvolume{100},
\bfpage{035306}
(\byear{2019}).
\doiurl{10.1103/PhysRevB.100.035306}
\end{barticle}
\endbibitem

%%% 53
\bibitem{Savona:SSComm93(1995)}
\begin{barticle}
\bauthor{\bsnm{Savona}, \binits{V.}},
\bauthor{\bsnm{Andreani}, \binits{L.C.}},
\bauthor{\bsnm{Schwendimann}, \binits{P.}},
\bauthor{\bsnm{Quattropani}, \binits{A.}}:
\batitle{Quantum well excitons in semiconductor microcavities: Unified treatment of weak and strong coupling regimes}.
\bjtitle{Solid State Communications}
\bvolume{93}(\bissue{9}),
\bfpage{733}--\blpage{739}
(\byear{1995}).
\doiurl{10.1016/0038-1098(94)00865-5}
\end{barticle}
\endbibitem

%%% 54
\bibitem{Combescot2008}
\begin{barticle}
\bauthor{\bsnm{Combescot}, \binits{M.}},
\bauthor{\bsnm{Betbeder-Matibet}, \binits{O.}},
\bauthor{\bsnm{Dubin}, \binits{F.}}:
\batitle{The many-body physics of composite bosons}.
\bjtitle{Physics Reports}
\bvolume{463}(\bissue{5-6}),
\bfpage{215}--\blpage{320}
(\byear{2008})
\end{barticle}
\endbibitem

%%% 55
\bibitem{Podolsky:PR34(1929)}
\begin{barticle}
\bauthor{\bsnm{Podolsky}, \binits{B.}},
\bauthor{\bsnm{Pauling}, \binits{L.}}:
\batitle{The momentum distribution in hydrogen-like atoms}.
\bjtitle{Phys. Rev.}
\bvolume{34},
\bfpage{109}--\blpage{116}
(\byear{1929}).
\doiurl{10.1103/PhysRev.34.109}
\end{barticle}
\endbibitem

%%% 56
\bibitem{Kavoulakis:PRB55(1997)}
\begin{barticle}
\bauthor{\bsnm{Kavoulakis}, \binits{G.M.}},
\bauthor{\bsnm{Chang}, \binits{Y.-C.}},
\bauthor{\bsnm{Baym}, \binits{G.}}:
\batitle{Fine structure of excitons in ${\mathrm{cu}}_{2}$o}.
\bjtitle{Phys. Rev. B}
\bvolume{55},
\bfpage{7593}--\blpage{7599}
(\byear{1997}).
\doiurl{10.1103/PhysRevB.55.7593}
\end{barticle}
\endbibitem

\end{thebibliography}

%Word count
%\newpage
%\detailtexcount{article}

%TC:endignore

\end{document}